\documentclass[12pt]{iopart}
\usepackage{epsfig,float,iopams,setstack}   

\DeclareMathAlphabet{\mathsfsl}{OT1}{cmss}{m}{sl}
\begin{document}

\title{Celestial mechanics in Kerr spacetime}

\author{W Schmidt\footnote{ Present address: Max-Planck-Institut
    f\"{u}r Astrophysik, Karl-Schwarzschild-Str. 1, Postfach 1317,
    D-85741 Garching, Germany}}

\address{Department of Physics and Astronomy, University of Wales, Cardiff, \\
  5 The Parade, Cardiff CF24 3YB, United Kingdom}

\ead{wolfram@mpa-garching.mpg.de}

\begin{abstract}
  The dynamical parameters conventionally used to specify the orbit of
  a test particle in Kerr spacetime are the energy $E$, the axial
  component of the angular momentum, $L_{z}$, and Carter's constant
  $Q$. These parameters are obtained by solving the Hamilton-Jacobi
  equation for the dynamical problem of geodesic motion. Employing the
  action-angle variable formalism, on the other hand, yields a
  different set of constants of motion, namely, the fundamental
  frequencies $\omega_{r}$, $\omega_{\theta}$ and $\omega_{\phi}$
  associated with the radial, polar and azimuthal components of
  orbital motion. These frequencies, naturally, determine the time
  scales of orbital motion and, furthermore, the instantaneous
  gravitational wave spectrum in the adiabatic approximation. In this
  article, it is shown that the fundamental frequencies are geometric
  invariants and explicit formulas in terms of quadratures are
  derived.  The numerical evaluation of these formulas in the case of
  a rapidly rotating black hole illustrates the behaviour of the
  fundamental frequencies as orbital parameters such as the semi-latus
  rectum $p$, the eccentricity $e$ or the inclination parameter
  $\theta_{-}$ are varied. The limiting cases of circular, equatorial
  and Keplerian motion are investigated as well and it is shown that
  known results are recovered from the general formulas.
\end{abstract}


\submitto{\CQG}

\section{Introduction}

In Newtonian theory, there are two basic methods of analysing the
motion of a test mass under the attraction of a central, spherically
symmetric gravitational field: Either Hamiltonian mechanics is applied
to solve the problem in a spherical polar coordinate system or, for
bound orbits, one may invoke a canonical transformation to
action-angle variables. In the latter case, solving the
\emph{Hamilton-Jacobi} differential equation yields the integrals of
motion and, in particular, \emph{Kepler's third law} which for
elliptical orbits, in geometric units, reads~\cite{Gold_i,CarrOst}
\begin{equation}
  \label{eq:kepler}
  M\Omega_{\mathrm{K}} = \left(\frac{1-e^2}{p}\right)^{3/2},
\end{equation}
where $\Omega_{\mathrm{K}}$ is the angular frequency of orbital
revolution, $e$ is the eccentricity of the orbit, $p/(1-e^{2})$ its
semi-major axis and $M$ is the mass which acts as the source of the
central field.

A generalisation of the Hamilton-Jacobi method to the relativistic
motion of a test mass in Kerr spacetime, as proposed by
Carter~\cite{Cart68}, shows that the motion is still separable and
there is a complete set of constants of motion: The energy $E$
as measured by an observer at spatial infinity, the axial angular
momentum component $L_{z}$ and the Carter constant $Q$.  Whereas $E$
and $L_{z}$ reflect isometries of the spacetime geometry, $Q$ arises
from the separation of the radial and polar components of motion. The
resulting first-order equations of motion can be used to compute the
particle's worldline and, in limiting cases such as circular and
equatorial orbits, to calculate periods of motion.

However, no attempt has been made yet to formulate a relativistic
generalisation of action-angle variables for geodesic motion in Kerr
spacetime and to calculate the dynamical frequencies of arbitrary
bound non-plunging orbits. Exactly these frequencies would be relevant
to the computation of gravitational radiation emitted by stellar-mass
compact objects in the vicinity of super-massive black holes if a
perturbation theory based on the adiabatic approximation is employed.
In this approximation one considers the motion as being nearly
geodesic under the assumption that the time scale of radiation
reaction, $T_{\mathrm{RR}}$, is much larger than the dynamical time
scales $T_{r}$, $T_{\theta}$ and $T_{\phi}$ of the three components of
orbital motion. Recently, Hughes~\cite{Hughes01} has raised the
question what time scales should, in general, be chosen for $T_{r}$,
$T_{\theta}$ and $T_{\phi}$, as there is no obvious way of calculating
orbital periods from the integrals of motion if the orbit is neither
circular nor equatorial.

In the following, it is shown that suitable action-angle variables can
be formulated within a fully relativistic framework for general orbits
in Kerr spacetime and formulas for dynamical frequencies in terms of
numerically solvable integrals are derived.

\section{Dynamical equations and integrals of motion}

\label{sct:eqn_intgr}

Let us first review the procedure applied by Carter~\cite{Cart68} in
order to determine the constants of motion.  We define the
relativistic Hamiltonian for geodesic motion of a particle that
possesses no spin in a spacetime with given metric in the absence of
an electromagnetic field as~\cite{MTW}
\begin{equation}
  \label{eq:hamilt}
  H(x^{\alpha},p_{\beta}) = \frac{1}{2}g^{\mu\nu}p_{\mu}p_{\nu}.
\end{equation}
In this equation, the metric components $g^{\mu\nu}$ are considered to
be functions of the coordinates $x^{\alpha}$ and the quantities
$p_{\beta}$ are the conjugate momenta of the particle associated with
these coordinates.  Substituting the contravariant components
$g^{\mu\nu}$ of the Kerr metric in the Boyer-Lindquist coordinate
representation~\cite{FroNovi}, the explicit form of the Hamiltonian is
given by
\begin{eqnarray}
  \label{eq:hamilt_bl}
\fl H^{(\mathrm{BL})}(x^{\alpha},p_{\beta}) = & 
  - \frac{(r^{2}+a^{2})^{2}-\Delta a^{2}\sin^{2}\theta}{2\Delta\Sigma}\,
  (p_{t})^{2} - \frac{2a M r}{\Delta\Sigma}\,p_{t}\,p_{\phi} \nonumber\\
  & +\frac{\Delta-a^{2}\sin^{2}\theta}{2\Delta\Sigma\sin^{2}\theta}\,
  (p_{\phi})^{2} + \frac{\Delta}{2\Sigma}\,(p_{r})^{2} 
  + \frac{1}{2\Sigma}\,(p_{\theta})^{2},
\end{eqnarray}
where $\Delta = r^{2} - 2M r + a^{2}$ and $\Sigma = r^{2} +
a^{2}\cos^{2}\theta$. 

A complete set of constants of motion can be determined if a canonical
transformation
$\Phi:(x^{\alpha},p_{\beta})\mapsto(X^{\alpha},P_{\beta})$ is found
such that the Hamiltonian becomes cyclic in all of the new generalised
coordinates $X^{\alpha}$, i.e.,
$(H^{(\mathrm{BL})}\circ\Phi^{-1})(X^{\alpha},P_{\beta})=
H^{(\mathrm{cycl})}(P_{\beta})$, and the transformed momenta
$P_{\beta}$ are thus conserved along the worldline of the particle.
The generator of such a canonical transformation,
$W(x^{\alpha},P_{\beta})$, is called the \emph{characteristic
  function} and is determined by the \emph{Hamilton-Jacobi}
differential equation
\begin{equation}
  g^{\mu\nu}\frac{\partial W}{\partial x^{\mu}}
  \frac{\partial W}{\partial x^{\nu}} + \mu^{2} = 0.
\end{equation}
Solving this equation, yields three constants of integration, $E$,
$L_{z}$ and $Q$, apart from $-\mu^{2}/2$, the value of the Hamiltonian
evaluated along the worldline of the particle. The transformed
conjugate momenta $P_{\beta}$ can be chosen as any functions of the
these constants, i.e., $P_{\beta}=f_{\beta}(-\mu^{2}/2,E,L_{z},Q)$,
assuming that $f$ is bijective and $\mathcal{C}^{\infty}$, and the
generalised coordinates $X^{\alpha}=\partial W/\partial P_{\alpha}$
are obtained from Hamilton's equations of motion,
\begin{equation}
  \label{eq:freq_gen}
  \mu\frac{\rmd X^{\alpha}}{\rmd \tau} =
  \frac{\partial H}{\partial P_{\alpha}}^{(\mathrm{cycl})}=\nu^{\alpha},
\end{equation}
where each $\nu^{\alpha}$ is a constant. The generalised coordinates
are therefore given by
$X^{\alpha}(\tau)=X^{\alpha}(0)+\nu^{\alpha}\tau$ if the Hamiltonian
is cyclic in all of these coordinates.

For geodesic motion in Kerr spacetime, the constants of motion
$E:=-p_{t}$ and $L_{z}:= p_{\phi}$ are related to the isometries of
the metric in the coordinates $t$ and $\phi$.  $E$ and $L_{z}$
can be interpreted, respectively, as the particle's energy and axial
angular momentum component as seen by an observer at spatial infinity.
The third constant of motion, $Q$, is obtained by separating radial
and polar motion and is called \emph{Carter's constant}. In terms of
$\mu$, $E$, $L_{z}$ and $Q$, the characteristic function $W$ is given
by
\begin{equation}
  W = -E t + \int\frac{\sqrt{R}}{\Delta}\,\rmd r
      + \int\sqrt{\Theta}\,\rmd \theta + L_{z}\phi,
\end{equation}
where
\begin{eqnarray}
  \label{eq:capt_r}
  R      & = \left[(r^2+a^2)E-a L_{z}\right]^2
             - \Delta[\mu^2 r^2+(L_{z}-a E)^2+Q], \\
  \label{eq:capt_theta}
  \Theta & = Q - \left[(\mu^2-E^2)a^2
             + \frac{L_{z}^2}{\sin^2\theta}\right]\cos^2\theta.
\end{eqnarray}
Since $p_{\beta}=g_{\beta\alpha}\partial W/\partial x^{\alpha}$, the
following set of first-order equations of motion is
obtained~\cite{Cart68,MTW}:
\begin{eqnarray}
  \label{eq:motion_t}
  \mu\Sigma\frac{\rmd t}{\rmd \tau}      & =
  \frac{r^{2}+a^{2}}{\Delta}P - a\left(a E\sin^{2}\theta-L_{z}\right), \\
  \label{eq:motion_r}
  \mu\Sigma\frac{\rmd r}{\rmd \tau}      & = \pm\sqrt{R}, \\
  \label{eq:motion_theta}
  \mu\Sigma\frac{\rmd \theta}{\rmd \tau} & = \pm\sqrt{\Theta}, \\
  \label{eq:motion_phi}
  \mu\Sigma\frac{\rmd \phi}{\rmd \tau}   & =
  \frac{a}{\Delta}P - a E + \frac{L_{z}}{\sin^{2}\theta},
\end{eqnarray}
where $P = E(r^{2}+a^{2}) - a L_{z}$.

If the conjugate momenta are chosen such that the Hamiltonian becomes
identically equal to $P_{0}$, i.e.,
$H^{(\mathrm{id})}(P_{\beta})=P_{0}$, then we obtain $\nu^{0}=1$ and
$\nu^{k}=0$. Substituting the characteristic function into the
identities $\partial W/\partial P_{0}=\tau+X^{0}(0)$ and $\partial
W/\partial P_{k}=X^{k}(0)$ and adjusting the constants $X^{\alpha}(0)$
in a suitable way, we find the following well-known integrals of
motion~\cite{Cart68,MTW}:
\begin{eqnarray}
  \label{eq:intgr_proper_time}
  \tau-\tau_{0} &= \int_{r_{0}}^{r}\frac{r'^2}{\sqrt{R}}\,\rmd r'
                   +\int_{\theta_{0}}^{\theta}
                   \frac{a^2\cos^2\theta'}{\sqrt{\Theta}}\,\rmd\theta',\\
  \label{eq:intgr_coord_time}
  t-t_{0}       &= \frac{1}{2}\int_{r_{0}}^{r}\frac{1}{\Delta\sqrt{R}}
                   \frac{\partial R}{\partial E}\,\rmd r'
                   +\frac{1}{2}\int_{\theta_{0}}^{\theta}\frac{1}{\sqrt{\Theta}}
                   \frac{\partial\Theta}{\partial E}\,\rmd\theta',\\
  \label{eq:intgr_angle}
  \phi-\phi_{0} &= -\frac{1}{2}\int_{r_{0}}^{r}\frac{1}{\Delta\sqrt{R}}
                   \frac{\partial R}{\partial L_{z}}\,\rmd r'
                   -\frac{1}{2}\int_{\theta_{0}}^{\theta}\frac{1}{\sqrt{\Theta}}
                   \frac{\partial\Theta}{\partial L_{z}}\,\rmd\theta',
\end{eqnarray}
and, finally,
\begin{equation}
  \label{eq:rad_pol}
  \int_{r_{0}}^{r}\frac{\rmd \theta'}{\sqrt{R}}=
  \int_{\theta_{0}}^{\theta}\frac{\rmd \theta'}{\sqrt{\Theta}}.
\end{equation}
In general, there are distinct turning points of the radial and polar
motions which are asynchronously passed by the particle because
$r(\tau)$ and $\theta(\tau)$ are not periodic functions of time. In
consequence, equation~(\ref{eq:intgr_coord_time}) evaluates to
different intervals of coordinate time when integrating over various
cycles of radial or polar motion and, for this reason, there is no
obvious way of calculating the time scales $T_{r}$, $T_{\theta}$ and
$T_{\phi}$ from the above integrals.

\section{The fundamental frequencies}
\label{sct:fund_freq}

Even though it might seem that there is no useful notion of orbital
frequencies if the coordinate functions representing the motion are
not periodic, it is still possible to find a representation in which
\emph{complete periodicity} becomes manifest. However, such a
representation is not found by means of a coordinate transformation in
the usual sense, but by virtue of the more general concept of a
\emph{canonical transformation} of both spacetime coordinates and
conjugate momenta. By its definition, a canonical transformation acts
on the cotangential bundle $T^{\ast}(\mathcal{M}_{\mathrm{Kerr}})$ of
Kerr spacetime, as opposed to a pure coordinate transformation between
charts on the basis manifold $\mathcal{M}_{\mathrm{Kerr}}$.  In
analogy to phase space trajectories in non-relativistic Hamiltonian
mechanics, let us consider the locus of all points in
$T^{\ast}(\mathcal{M}_{\mathrm{Kerr}})$ which are passed by a particle
in the course of its motion along a particular orbit.  We shall denote
the image of that set under the coordinate chart corresponding to
Boyer-Lindquist coordinates $x^{\alpha}$ and their conjugate momenta
$p_{\beta}$ by $\mathcal{T}_{\mu,E,L_{z},Q}$ for a given set of
constants of motion. 

The basic topological properties of $\mathcal{T}_{\mu,E,L_{z},Q}$ in
the case of a bound orbit of the first kind can be readily inferred
from the equations determining the conjugate momenta,
\begin{eqnarray}
  \label{eq:conj_momenta}
  \eqalign{
    p_{t}               = -E, \qquad & p_{\phi}       = L_{z}, \\
    \Delta^{2}p_{r}^{2} =  R, \qquad & p_{\theta}^{2} = \Theta.
  }
\end{eqnarray}
Firstly, since the orbital revolution progresses continuously in time,
$\mathcal{T}_{\mu,E,L_{z},Q}$ is not compact in timelike directions.
Secondly, the motion is bound both in the $r$- and in the
$\theta$-domain. Moreover, since $p_{r}$ is solely a function of $r$
and, likewise, $p_{\theta}$ is a function of $\theta$ only, both
radial and polar motion are of compact support and of the libration
type provided that the orbit is \emph{bound} and \emph{stable}.
Otherwise, the orbit either plunges and, hence, the particle
approaches the horizon towards future timelike infinity or, if the
orbit is only marginally stable, it asymptotically approached the
inner turning point. Thirdly, the azimuthal motion is a rotation and,
of course, it is of compact support as well because $\phi=0$ is to be
identified with $\phi=2\pi$.

According to a theorem by Arnold~\cite{Thirr_i}, that was formulated
in the framework of non-relativistic mechanics, the phase space
trajectory of a \emph{fully integrable} system of $N$ degrees of
freedom is diffeomorphic to the $N$-torus
$\mathcal{T}^{N}=(\mathcal{S}_{1})^{N}$ whenever the motion of the
system is of \emph{compact support}.  It seems plausible, that this
theorem is applicable to relativistic problems as well if we account
for the distinct role of the timelike coordinate which is related to
the non-compactness of the worldline of a particle.  Accordingly, as a
manifestation of the weak equivalence principle, geodesic motion in
Kerr spacetime has only three \emph{physical} degrees of freedom
because orbits are the same regardless of the mass $\mu$ of the
particle.  Since $\mathcal{T}_{\mu,E,L_{z},Q}$ is defined with respect
to a particular coordinate system, there is a well defined notion of
projecting this set in the directions of the pair of conjugate
variables $(t,p_{t})$.  The resulting projected set,
$\hat{\mathcal{T}}_{\mu,E,L_{z},Q}$, resides in the six-dimensional
space spanned by the spacelike conjugate pairs $(r,p_{r})$,
$(\theta,p_{\theta})$ and $(\phi,p_{\phi})$ and corresponds to a phase
space. Indeed, from the topological considerations in the previous
paragraph, we see that $\hat{\mathcal{T}}_{\mu,E,L_{z},Q}$ is
diffeomorphic to the three-torus $\mathcal{T}^{3}$.

Utilising this toroidal topology, let us now define for each spatial
component a closed curve
$\mathcal{C}_{k}\subset\,\hat{\mathcal{T}}_{\mu,E,L_{z},Q}$ which
circumscribes the $k$-th torus and can be contracted into a single
point with respect to the other tori. If $\psi$ is the coordinate
chart on the cotangential bundle producing the Boyer-Lindquist
coordinates and its conjugate momenta, then
$\psi^{-1}(\mathcal{C}_{k})$ is the pre-image of the $k$-th loop in
$T^{\ast}(\mathcal{M}_{\mathrm{Kerr}})$.  Furthermore, if we note that
the \emph{canonical one-form} of Hamiltonian mechanics,
$\boldsymbol{\Theta}=-E\,\boldsymbol{\rmd}t+p_{r}\boldsymbol{\rmd}r+
p_{\theta}\boldsymbol{\rmd}\theta+L_{z}\boldsymbol{\rmd}\phi$, is the
one-form corresponding to the relativistic four-momentum $\mathbf{p}$,
i.e.,
\begin{equation}
  \langle\boldsymbol{\Theta},\bi{p}\rangle = p_{\alpha}p^{\alpha} = -\mu^{2},
\end{equation}
then we can define an \emph{action variable} $J_{k}$ for the $k$-th
spatial component as
\begin{equation}
  \label{eq:action_var}
  J_{k} = \frac{1}{2\pi}\oint_{\psi^{-1}(\mathcal{C}_{k})}\!\boldsymbol{\Theta}.
\end{equation}
It might appear that such an action variable ultimately depends on the
choice of the closed curve $\mathcal{C}_{k}$ which is defined with
respect to a particular coordinate system and, thus, one would assume
that it is not a geometric invariant. However, $J_{k}$ is independent
of the shape of $\mathcal{C}_{k}$ because $\boldsymbol{\Theta}$ is
\emph{closed} on $\psi^{-1}(\mathcal{T}_{\mu,E,L_{z},Q})$, i.e.,
$\boldsymbol{\rmd\Theta}=0$. The closedness of the canonical one-form
is shown by substituting the expressions~(\ref{eq:conj_momenta}) for
the conjugate momenta and applying the rules of exterior
differentiation\footnote{ It is actually the separability of the
  components of motion which renders the canonical one-form being
  closed on the dynamical submanifold in the cotangential bundle.  }.
The generalised Stokes theorem then yields the desired property of
$J_{k}$~\cite{Thirr_ii}.  Furthermore, since transformations between
coordinate charts preserve topological features, we could have defined
the curves $\mathcal{C}_{k}$ with respect to any coordinate system and
then found their pre-images in
$T^{\ast}(\mathcal{M}_{\mathrm{Kerr}})$. It is only the topological
features of $\mathcal{C}_{k}$ with respect to the three-torus that
matters! Therefore we conclude that the action
variables~(\ref{eq:action_var}) are, indeed, geometric invariants.
For convenience, however, we shall denote these variables as $J_{r}$,
$J_{\theta}$ and $J_{\phi}$ as if they were specific for the
Boyer-Lindquist coordinate system.

Since motion in Kerr spacetime is separable in the coordinates $r$,
$\theta$ and $\phi$, we may choose each $\mathcal{C}_{k}$ such that it
is located completely within the $(x^{k},p_{k})$-plane. Then the
action variables can be easily calculated from cyclic integrals over
the spatial conjugate momenta in the Boyer-Lindquist coordinate
representation:
\begin{eqnarray}
  \label{eq:action_var_r}
  J_{r} & = \frac{1}{2\pi}\oint p_{r}\,\rmd r =
  \frac{1}{2\pi}\oint\frac{\sqrt{R}}{\Delta}\,\rmd r, \\
  \label{eq:action_var_theta}
  J_{\theta} & = \frac{1}{2\pi}\oint
  p_{\theta}\,\rmd \theta =
  \frac{1}{2\pi}\oint\Theta\,\rmd \theta, \\
  \label{eq:action_var_phi}
  J_{\phi}   & = \frac{1}{2\pi}\oint p_{\phi}\,\rmd \phi = L_{z}.
\end{eqnarray}
As can be seen from the above expressions, $J_{r}$, $J_{\theta}$ and
$J_{\phi}$ are functions of the mass $\mu$ and the three constants of
motion $E$, $L_{z}$ and $Q$.

The generator of the transformation to the action variables $J_{k}$ is
given by $W=\int\boldsymbol{\Theta}$. It is not globally defined, but
changes by an amount of $2\pi J_{k}$ when integrating over the path
$\mathcal{C}_{k}$.  As a consequence, if the spatial coordinate
$x^{k}$ goes through $N$ complete cycles and the other coordinates are
kept unchanged, then the generalised coordinate $w^{k}=\partial
W/\partial J_{k}$ associated with $J_{k}$ changes by $2\pi
N$.\footnote{ Such a change has to be understood as a virtual
  displacement. } This result implies that $w_{k}$ is a phase or
\emph{angle variable} and, since
${\scriptstyle\Delta}w^{k}=\omega_{k}{\scriptstyle\Delta}\tau$, the
constant $\omega_{k}$ can be interpreted as a frequency. Moreover, as
a consequence of the geometrically invariant definition of the action
variables $J_{k}$, the frequencies $\omega_{k}$ specify fundamental
properties of the orbital motion independent of any particular
coordinate representation. Therefore, they are called the
\emph{fundamental frequencies} of the system. However, these
frequencies are subject to a straightforward physical interpretation
only with respect to the spatial coordinates $r$, $\theta$ and $\phi$,
when the paths $\mathcal{C}_{k}$ are correspondingly chosen as cycles of,
respectively, radial, polar and azimuthal motion. Accordingly, we
shall use the notation $\omega_{r}$, $\omega_{\theta}$ and
$\omega_{\phi}$ for the fundamental frequencies of an orbit in Kerr
spacetime.

The standard procedure of determining fundamental frequencies is to
find the explicit form of the Hamiltonian in the action-angle
representation, $H^{(\mathrm{aa})}$, and to calculate the frequencies
from the partial derivatives with respect to the action variables
$J_{k}$~\cite{Gold_i}:
\begin{equation}
  \mu\omega_{k} = \frac{\partial H}{\partial J_{k}}^{(\mathrm{aa})}.
\end{equation}
Unfortunately, the definition of the radial and polar action variables
in terms of the non-trivial integrals~(\ref{eq:action_var_r})
and~(\ref{eq:action_var_theta}) does not admit an explicit inversion
because none of these integrals can be solved analytically. However,
as is shown in \ref{sct:determ_freq}, the derivatives
$\partial H^{(\mathrm{aa})}/\partial J_{k}$ and, hence, the
frequencies $\omega_{k}$ can be found even without knowing the
functional form of the Hamiltonian $H^{(\mathrm{aa})}$ if the
theorem on implicit functions is employed.  This procedure results in
equations~(\ref{eq:h_derv_jr})--(\ref{eq:h_derv_jphi}) of
\ref{sct:determ_freq} and the fundamental frequencies are
therefore given by
\begin{eqnarray}
  \frac{2\pi}{\mu\omega^{r}} & = 
  2 a^{2}z_{+}^2\left[1-\frac{E(k)}{K(k)}\right]X(r_{1},r_{2}) + 
  2 Y(r_{1},r_{2}), \\
  \frac{2\pi}{\mu\omega^{\theta}} & = 
  \frac{4 a^{2} z_{+}}{\beta}\left[K(k)-E(k)\right] +
  \frac{4}{\beta z_{+}}\frac{X(r_{1},r_{2})}{Y(r_{1},r_{2})}K(k), \\
  \frac{2\pi}{\mu\omega^{\phi}} & = 
  2\pi\frac{a^{2}z_{+}^{2}[K(k)-E(k)]X(r_{1},r_{2}) + K(k)Y(r_{1},r_{2})
            }{L_{z}[\Pi(z_{-}^{2},k)-K(k)]X(r_{1},r_{2}) +
  K(k)Z(r_{1},r_{2})},
\end{eqnarray}
where $r_{1}$ and $r_{2}$ are the turning points of radial motion,
$z_{\pm}^{2}$ are the two roots of the equation $\Theta(z)=0$ when
substituting $\cos\theta=z$ in $\Theta$, $k=z_{-}^{2}/z_{+}^{2}$,
$\beta^{2}=a^{2}(\mu^{2}-E^{2})$, $K(k)$, $E(k)$ and
$\Pi(z_{-}^{2},k)$ are, respectively, the complete elliptical
integrals of the first, second and third kind,
\begin{eqnarray}
  \label{eq:ell_integr_k}
  K(k) & =\int_{0}^{\pi/2}\frac{\rmd \psi}{\sqrt{1-k\sin^{2}\psi}}, \\
  \label{eq:ell_integr_e}
  E(k) & =\int_{0}^{\pi/2}\sqrt{1-k\sin^{2}\psi}\,\rmd \psi, \\
  \label{eq:ell_integr_pi}
  \Pi({z_{-}^{2},k}) & =
  \int_{0}^{\pi/2}\frac{\rmd\psi}{\left(1-z_{-}^{2}\sin^{2}\psi\right)
  \sqrt{1-k\sin^{2}\psi}},
\end{eqnarray}
and $X(r_{1},r_{2})$, $Y(r_{1},r_{2})$ and $Z(r_{1},r_{2})$ are radial integrals 
defined by
\begin{eqnarray}
  \label{eq:rad_integr_x}
  X(r_{1},r_{2}) & = \int_{r_{1}}^{r_{2}}\frac{\rmd r}{\sqrt{R}}, \\
  \label{eq:rad_integr_y}
  Y(r_{1},r_{2}) & = \int_{r_{1}}^{r_{2}}\frac{r^{2}\,\rmd r}{\sqrt{R}}, \\
  \label{eq:rad_integr_z}
  Z(r_{1},r_{2}) & = \int_{r_{1}}^{r_{2}}
  \frac{r\left[L_{z}r-2M(L_{z}-a E)\right]}{\Delta\sqrt{R}}\,\rmd r.
\end{eqnarray}

The above equations for the fundamental frequencies, however, have two
deficiencies: Firstly, they are suggestive of $\omega_{k}$ being
dependent on the mass $\mu$ of the particle which is, of course, not
the case. For this reason, we will cast these equations into a
dimensionless scale-invariant form, in which $\mu$ does not appear
anymore. Secondly, the radial functions $X$, $Y$ and $Z$ are not
proper integrals because the integrated functions are divergent at the
turning points $r_{1}$ and $r_{2}$. Fortunately, they are transformed
into well-behaved integrals by virtue of the substitution
$r=M\tilde{r}=p M/(1+e\cos\chi)$, where $\chi$ varies from $0$ to
$2\pi$ as $r$ goes through a complete cycle. Although this
substitution originates from the pure Kepler ellipse, it is actually
still expedient even for relativistic orbits that may be very
different in shape. In analogy to non-relativistic orbits, $p$ is
called the \emph{semi-latus rectum} and $e$ is the \emph{eccentricity}
of the orbit. Working out this substitution, yields the following
equations for the dimensionless fundamental frequencies
$\tilde{\omega}_{k}=M\omega_{k}$:
\begin{eqnarray}
  \label{eq:freq_r}
  \tilde{\omega}_{r} & = \frac{\pi p K(k)}{(1-e^{2})\Lambda}, \\
  \label{eq:freq_theta}
  \tilde{\omega}_{\theta} & = \frac{\pi\tilde{\beta}z_{+}\tilde{X}}{2\Lambda}, \\
  \label{eq:freq_phi}
  \tilde{\omega}_{\phi} & = \frac{(\tilde{Z}-\tilde{L}_{z}\tilde{X})K(k)
                            + \tilde{L}_{z}\tilde{X}\Pi(z_{-}^{2},k)}{\Lambda},
\end{eqnarray}
where the tilde on top of each symbol indicates the corresponding
dimensionless quantity (see \ref{sct:const_of_motion}) and
the constant $\Lambda$ is defined by
\begin{equation}
  \Lambda = (\tilde{Y}+\tilde{a}^{2}z_{+}^{2}\tilde{X})K(k) - 
  \tilde{a}^{2}z_{+}^{2}\tilde{X}E(k).
\end{equation}
The dimensionless radial integrals in the $\chi$-representation are given by

\begin{eqnarray}
  \label{eq:integr_x_diml}
  \tilde{X} & = \mu M\frac{p}{1-e^{2}}\,X(r_{1},r_{2}) =
  \int_{0}^{\pi}\frac{\rmd \chi}{\sqrt{J(\chi)}}, \\
  \label{eq:integr_y_diml}
  \tilde{Y} & = \frac{\mu}{M}\cdot\frac{p}{1-e^{2}}\,Y(r_{1},r_{2}) =
  \int_{0}^{\pi}\frac{p^{2}\,\rmd \chi}{(1+e\cos\chi)^{2}\sqrt{J(\chi)}}, \\
  \label{eq:integr_z_diml}
  \tilde{Z} & = \frac{p}{1-e^{2}}\,Z(r_{1},r_{2}) =
  \int_{0}^{\pi}\frac{G(\chi)\,\rmd \chi}{H(\chi)\sqrt{J(\chi)}},
\end{eqnarray}
where the functions $G$, $H$ and $J$ are defined by
\begin{eqnarray}
  \label{eq:def_j}
  J(\chi) = & (1-\tilde{E}^2)(1-e^{2}) 
                + 2\left(1-\tilde{E}^{2}-\frac{1-e^{2}}{p}\right)
                  (1+e\cos\chi) \nonumber\\
              & + \left\{(1-\tilde{E}^2)\frac{3+e^{2}}{1-e^{2}}-\frac{4}{p}+
                         \left[\tilde{a}^{2}(1-\tilde{E}^{2})+
                               \tilde{L}_{z}^{2}+
                               \tilde{Q}\right]\frac{1-e^{2}}{p^{2}}
                  \right\} \nonumber\\
              & \times (1+e\cos\chi)^{2}, \\
  \label{eq:def_h}
  H(\chi) = & 1 - \frac{2}{p}(1+e\cos\chi)
                + \frac{\tilde{a}^{2}}{p^{2}}(1+e\cos\chi)^{2}, \\
  \label{eq:def_g}
  G(\chi) = & \tilde{L_{z}}
                - \frac{2(\tilde{L}_{z}-\tilde{a}\tilde{E})}{p}(1+e\cos\chi).
\end{eqnarray}
In the above equations, the constants of motion $E$, $L_{z}$ and $Q$
are understood as functions of the orbital parameters $p$, $e$ and
$z_{-}=\cos\theta_{-}$. Although analytical expressions for these
functions can be derived, it is hardly feasible to substitute these
expressions for $E$, $L_{z}$ and $Q$. For given orbital parameters, it
is much easier to calculate the constants of motion numerically, as
explained in \ref{sct:const_of_motion}, and to substitute the
resulting values into $G(\chi)$, $H(\chi)$ and $J(\chi)$.

Apart from the three frequencies $\omega_{r}$, $\omega_{\theta}$ and
$\omega_{\phi}$, a fourth constant of motion is obtained,
which is associated with the timelike generalised coordinate
$X^{0}_{(\mathrm{aa})}(\tau)=X^{0}_{(\mathrm{aa})}(0)+\gamma\tau$ in
the action-angle variable representation.  According to
equation~(\ref{eq:h_derv_p0}) of \ref{sct:determ_freq}, this
constant is given by
\begin{equation}
  \mu\gamma = 
  \frac{\left[W(r_{1},r_{2})+a^{2}z_{+}^{2}E X(r_{1},r_{2})\right]K(k)-
        a^{2}z_{+}^{2}E X(r_{1},r_{2})E(k)
        }{\left[Y(r_{1},r_{2})+a^{2}z_{+}^{2}X(r_{1},r_{2})\right]K(k)- 
          a^{2}z_{+}^{2}X(r_{1},r_{2})E(k)},
\end{equation}
where the function $W(r_{1},r_{2})$ is defined by
\begin{equation}
  \label{eq:rad_integr_w}
  W(r_{1},r_{2}) = 
  \int_{r_{1}}^{r_{2}}\frac{r\left[r^{3}+a^{2}E r-2a(L_{z}-a E)\right]\,\rmd r
                            }{\Delta\sqrt{R}}.
\end{equation}
In dimensionless form, $\gamma$ can be expressed as
\begin{equation}
  \label{eq:gamma}
  \gamma = \frac{1}{\Lambda}
           \left[(\tilde{W}+\tilde{a}^{2}z_{+}^{2}\tilde{E}\tilde{X})K(k)- 
                 \tilde{a}^{2}z_{+}^{2}\tilde{E}\tilde{X}E(k)\right],
\end{equation}
where
\begin{equation}
  \label{eq:integr_w_diml}
  \tilde{W} = \frac{1}{M}\cdot\frac{p}{1-e^{2}}\,W(r_{1},r_{2}) =
  \int_{0}^{\pi}\frac{p^{2}F(\chi)\,\rmd \chi
                      }{(1+e\cos\chi)^{2}H(\chi)\sqrt{J(\chi)}},
\end{equation}
and the function $F$ is defined by
\begin{equation}
  \label{eq:def_f}
  F(\chi) = \tilde{E} + \frac{\tilde{a}^{2}\tilde{E}}{p^{2}}(1+e\cos\chi)^{2}
            - \frac{2\tilde{a}(\tilde{L}_{z}
            - \tilde{a}\tilde{E})}{p^{3}}(1+e\cos\chi)^{3}.
\end{equation}
As for the physical interpretation of $\gamma$, we note that the
conjugate momentum associated with the generalised coordinate
$X^{0}_{(\mathrm{aa})}$ is $P_{0}^{(\mathrm{aa})}=p_{t}=-E$, i.e., the
same as in the Boyer-Lindquist representation. For this reason, 
one would expect that $X^{0}_{(\mathrm{aa})}$ is in some sense an interval of
coordinate time. However, it cannot be equal to $t-t_{0}$ as
equation~(\ref{eq:intgr_coord_time}) tells us that the relation
between the elapse of proper time $\tau$ and the corresponding
interval of coordinate time, $t-t_{0}$, is not linear. On the other
hand, there are three particular scales of proper time, $\tau_{r}$,
$\tau_{\theta}$ and $\tau_{\phi}$, associated with the three
fundamental frequencies, $\omega_{r}$, $\omega_{\theta}$ and
$\omega_{\phi}$. Substituting each of these constants into the
expression for $X^{0}_{(\mathrm{aa})}$, three scales of
\emph{coordinate time} are produced:
\begin{equation}
  \label{eq:coord_time_scl}
  T_{k} = 
  X^{0}_{(\mathrm{aa})}(\tau_{k})-X^{0}_{(\mathrm{aa})}(0) =
  \frac{2\pi\gamma}{\omega_{k}}.
\end{equation}
This seems to be a natural definition of the coordinate-time scales
associated with the fundamental frequencies. As we shall see in
following Section, the above definition, indeed, reproduces the known
periods of coordinate time in the limits of, respectively, circular
and equatorial motion. Hence, we are led to the conjecture that
$\gamma$ can be interpreted as a \emph{gravitational Lorentz factor}
which relates the three dynamical scales of proper time, $\tau_{r}$,
$\tau_{\theta}$ and $\tau_{\phi}$, to the corresponding intervals of
coordinate time, $T_{r}$, $T_{\theta}$ and $T_{\phi}$.

In the general case, this conjecture has the following implication:
The ratio of time scales associated with two particular components of
motion, say, radial and polar motion, is the same regardless of the
frame of reference in which these time scales are defined. For
example, if an observer at infinity saw the intervals of time $T_{r}$
and $T_{\theta}$, he would find that the ratio $T_{r}/T_{\theta}$ is equal
to $\tau_{r}/\tau_{\theta}$. Of course, this is exactly what one would
expect, provided that the fundamental frequencies truly arise from
invariant properties of orbital motion. In conclusion, the ratio of
time scales should be independent of the frame of reference in use.
With respect to coordinate time $t$, this is inherently ensured by the
above definition of $T_{r}$, $T_{\theta}$ and $T_{\phi}$.

Ultimately, the significance of the fundamental frequencies lies in the
possibility of expanding a dynamical quantity $q(\tau)$ in a Fourier
series with harmonics
$\omega_{klm}=k\omega_{r}+l\omega_{\theta}+m\omega_{\phi}$ of these
frequencies~\cite{Gold_i}:
\begin{equation}
  q(\tau) = \sum_{k,l,m}q_{klm}\,\exp(\rmi\omega_{klm}\tau) =
  \sum_{\vec{k}}q_{\vec{k}}\,\exp(\rmi\vec{k}\cdot\vec{w}),
\end{equation}
where $\vec{k}=(k,l,m)$ and $\vec{k}\cdot\vec{w}=k w^{r}+l
w^{\theta}+m w^{\phi}= \omega_{klm}\tau$ according to definition of
the action variables $w^{r}$, $w^{\theta}$ and $w^{\phi}$. The Fourier
coefficients $q_{\vec{k}}$ of the above expansion are given by
\begin{equation}
\fl q_{\vec{k}} = \frac{1}{(2\pi)^{3}}
                  \int_{0}^{2\pi}\rmd w^{r}\,
                  \int_{0}^{2\pi}\rmd w^{\theta}\,
                  \int_{0}^{2\pi}\rmd w^{\phi}\,
                  q(\vec{k}\cdot\vec{w}/\omega_{\vec{k}})
                  \exp(-\rmi\vec{k}\cdot\vec{w}).
\end{equation}
Therefore, dynamical properties of the particle are completely
specified by discrete sets of numbers such as
$\{q_{\vec{k}}\}_{\vec{k}\in\mathbb{Z}^{3}}$. This is the very essence
of bound geodesic orbital motion in a Kerr spacetime being
\emph{quasi-periodic} rather than quasi-chaotic, as was proposed due
to the ergodic properties of these orbits.

\section{Limiting cases and numerical results}

As an important test to support the validity of the fundamental
frequencies derived in the previous Section, we will consider two
special cases of bound motion in Kerr spacetime: Firstly, the
limit of vanishing eccentricity and, secondly, purely equatorial
orbits.  In particular, we will see that previously derived equations
for the periods and the corresponding azimuthal angles of circular
orbits~\cite{Hughes00} can be reproduced from the general formulas.
Briefly summarised, the features of orbits in the two limiting cases
are as follows:
\begin{itemize}
\item Circular orbits ($e$ = 0): In Schwarzschild spacetime, Kepler's
  third law~(\ref{eq:kepler}) holds exactly for the period of
  azimuthal revolution with respect to coordinate time, i.e.,
  $T_{\phi}=2\pi M\tilde{r}_{0}^{3/2}$. For circular
  orbits in Kerr spacetime, there are two distinct frequencies
  depending on whether the particle co-revolves or counter-revolves
  with the rotation of the black hole. If the motion is not confined
  to the equatorial plane, then the particle changes periodically its
  altitude such that $2\pi/\omega_{\theta}$ is twice the elapse of
  proper time between two subsequent passages of the particle through
  the equatorial plane. The frequencies $\omega_{\theta}$ and
  $\omega_{\phi}$ are generally incommensurate. As a consequence, the
  azimuthal angle $\Phi_{\theta}=2\pi\omega_{\phi}/\omega_{\theta}$
  that is accumulated as the particle completes a cycle of polar
  motion is different from $2\pi$.
\item Equatorial orbits ($Q=0$): As a consequence of the different
  shape of the potential of radial motion as compared to the Newtonian
  potential, non-circular orbits do not close on themselves as the
  particle revolves around the gravitational centre. The total
  azimuthal angle $\Phi_{r}$ swept out by the particle as it moves
  from, say, the apastron at radius $r_{\mathrm{a}}=p/(1-e)$ to the
  periastron at radius $r_{\mathrm{p}}=p/(1+e)$ and back to the
  apastron is given by $\Phi_{r}=2\pi\omega_{\phi}/\omega_{r}$.
\end{itemize}
Actually, there is another limit, called polar orbits, for which
$L_{z}=0$. In this case, the particle crosses the spin axis of the
black hole and the inclination parameter is $z_{-}=1$. However, it
seems unlikely that polar orbits are astrophysically relevant and we
only refer to \cite{StoTsou87} for a general discussion.

\subsection{Circular orbits}

Orbits of zero eccentricity, $e=0$, are called circular. In
this case, the particle moves on a spherical surface of constant
radial coordinate, $r(\tau)=p M=r_{0}$, within the bounds of the
inclination $\theta$ given by
$\theta_{-}\leq\theta\leq\pi-\theta_{-}$. Since the perihelion and
aphelion radii coincide, we have $r_{\mathrm{p}}=r_{\mathrm{a}}=r_{0}$ and
$R(r_{0})=0$. In consequence, the integrals given by
equations~(\ref{eq:rad_integr_x})--(\ref{eq:rad_integr_z})
and~(\ref{eq:rad_integr_w}) are undefined in the limit of vanishing
eccentricity.  The corresponding quantities in the
$\chi$-representation, however, have well-defined limits because the
functions $F$, $G$, $H$ and $J$ are constant if $e=0$.  Hence, we
define
\begin{eqnarray}
  \eqalign{
    X(r_{0},r_{0}) = \frac{\pi}{\mu r_{0}J_{\mathrm{circ}}^{1/2}}, \qquad
  & W(r_{0},r_{0}) = \frac{\pi r_{0}F_{\mathrm{circ}}
                           }{H_{\mathrm{circ}}J_{\mathrm{circ}}^{1/2}}, \\
    Y(r_{0},r_{0}) = \frac{\pi r_{0}}{\mu J_{\mathrm{circ}}^{1/2}} , \qquad
  & Z(r_{0},r_{0}) = \frac{\pi M G_{\mathrm{circ}}
                           }{r_{0}H_{\mathrm{circ}}J_{\mathrm{circ}}^{1/2}},
  }
\end{eqnarray}
where the constants $F_{\mathrm{circ}}$, $G_{\mathrm{circ}}$,
$H_{\mathrm{circ}}$ and $J_{\mathrm{circ}}$ are obtained by
substituting $e=0$ in equations~(\ref{eq:def_j})--(\ref{eq:def_g})
and~(\ref{eq:def_f}).  The frequencies of the angular components of
motion are therefore given by
\begin{eqnarray}
  \label{eq:freq_r_circ}
  \tilde{\omega}_{r}^{(\mathrm{circ})} & =
  \frac{\tilde{r}_{0}J_{\mathrm{circ}}^{1/2}K(k)
        }{(\tilde{r}_{0}^{2}+\tilde{a}^{2}z_{+}^{2})K(k) - 
          \tilde{a}^{2}z_{+}^{2}E(k)}, \\
  \label{eq:freq_theta_circ}
  \tilde{\omega}_{\theta}^{(\mathrm{circ})} & =
  \frac{\pi\tilde{\beta}z_{+}
        }{2[(\tilde{r}_{0}^{2}+\tilde{a}^{2}z_{+}^{2})K(k) - 
          \tilde{a}^{2}z_{+}^{2}E(k)]}, \\
  \label{eq:freq_phi_circ}
  \tilde{\omega}_{\phi}^{(\mathrm{circ})} & =
  \frac{\tilde{a}(2\tilde{E}\tilde{r}_{0} -
        \tilde{a}\tilde{L}_{z})\tilde{\Delta}_{0}^{-1}K(k) +
        \tilde{L}_{z}\Pi(z_{-}^{2},k)
        }{(\tilde{r}_{0}^{2}+\tilde{a}^{2}z_{+}^{2})K(k) - 
          \tilde{a}^{2}z_{+}^{2}E(k)}.
\end{eqnarray}

The existence of a distinct radial frequency even in the case of
circular motion indicates that the limit of the integrated azimuthal
angle over a cycle of \emph{radial} motion, $\Phi_{r}$, does not
become $2\pi$ as the eccentricity approaches zero. Of course, this is
well known from equatorial orbits. On the other hand, we can calculate
the integrated azimuthal angle $\Phi_{\theta}$ as the particle goes
through one cycle of \emph{polar} motion by combining the equations of
motion~(\ref{eq:motion_theta}) and~(\ref{eq:motion_phi}). The
result is
\begin{equation}
  \label{eq:intgr_angl_circ}
  \Phi_{\theta} = 
  \frac{4}{\tilde{\beta}z_{+}}
  \left[\frac{\tilde{a}(2\tilde{E}\tilde{r}_{0}-\tilde{a}\tilde{L}_{z})
              }{\tilde{\Delta}_{0}}K(k) + \tilde{L}_{z}\Pi(z_{-}^{2},k)
  \right],
\end{equation}
where
$\tilde{\Delta}_{0}=\tilde{r}_{0}^{2}-2\tilde{r}_{0}+\tilde{a}^{2}$.
When comparing this expression for $\Phi_{\theta}$ with the
quotient of $\tilde{\omega}_{\phi}^{(\mathrm{circ})}$ and
$\tilde{\omega}_{\theta}^{(\mathrm{circ})}$, we see that the following
well known equation holds:
\begin{equation}
  \Phi_{\theta} = 2\pi\frac{\tilde{\omega}_{\phi}^{(\mathrm{circ})}
                            }{\tilde{\omega}_{\theta}^{(\mathrm{circ})}}.
\end{equation}
Furthermore, calculating the period of coordinate time $T_{\theta}$
for a cycle of polar motion from equation~(\ref{eq:intgr_coord_time}),
yields
\begin{equation}
\fl T_{\theta} = 
  \frac{4}{\tilde{\beta}z_{+}}
  \left\{\frac{\tilde{E}(\tilde{r}_{0}^{2} + \tilde{a}^{2})\tilde{r}_{0}^{2} -
               2\tilde{a}(\tilde{L}_{z}-\tilde{a}\tilde{E})\tilde{r}_{0}
               }{\tilde{\Delta}_{0}}K(k) + 
         \tilde{a}^{2}z_{+}^{2}\tilde{E}[K(k)-E(k)]
  \right\}.
\end{equation}
If the constant $\gamma$ given by equation~(\ref{eq:gamma}) is also
calculated in the limit $e=0$, we find that
\begin{equation}
  T_{\theta} = \frac{2\pi\gamma^{(\mathrm{circ})}
                     }{\omega_{\theta}^{(\mathrm{circ})}},
\end{equation}
which suggests that $\gamma^{(\mathrm{circ})}$ can be interpreted as the \emph{Lorentz
  factor} of circular motion in Kerr spacetime.

\subsection{Equatorial orbits}
\label{sct:equat}

If $Q=0$, then the equation for the turning points of polar motion,
$\Theta(z)=0$, admits the solutions $z_{-}^{2}=0$ and
$z_{+}^{2}=1+L_{z}^{2}/\beta^{2}>1$.  As a consequence, the motion of
the particle is confined to the equatorial plane, i.e.,
$\theta(\tau)=\pi/2$. Since $k=(z_{-}/z_{+})^{2}=0$, the elliptical
integrals become $K(0)=E(0)=\Pi(0,0)=\pi/2$. The resulting expressions
for the fundamental frequencies of equatorial orbits are thus given by
\begin{eqnarray}
  \omega_{r}^{(\mathrm{eq})} & = 
  \frac{\pi p}{(1-e^{2})\tilde{Y}_{\mathrm{eq}}}, \\
  \omega_{\theta}^{(\mathrm{eq})} & = 
  \tilde{\beta}z_{+}\,\frac{\tilde{X}_{\mathrm{eq}}}{\tilde{Y}_{\mathrm{eq}}}, \\
  \omega_{\phi}^{(\mathrm{eq})} & = 
  \frac{\tilde{Z}_{\mathrm{eq}}}{\tilde{Y}_{\mathrm{eq}}}.
\end{eqnarray}
As in the case of circular orbits, we still have three different frequencies.

Combining the equations of motion~(\ref{eq:motion_r})
and~(\ref{eq:motion_phi}), we can calculate the integrated azimuthal
angle $\Phi_{r}$ between two successive passages through, say, the
periastron and we find that
\begin{equation}
  \Phi_{r} = 2 Z_{\mathrm{eq}} = \frac{2(1-e^{2})}{p}\tilde{Z}_{\mathrm{eq}}.
\end{equation}
According to equation~(\ref{eq:intgr_coord_time}), the corresponding
period of coordinate time is given by $T_{r} =2 W_{\mathrm{eq}}$.
Using
$\gamma^{(\mathrm{eq})}=\tilde{W}_{\mathrm{eq}}/\tilde{Y}_{\mathrm{eq}}$,
we obtain
\begin{equation}
  T_{r} = \frac{2\pi\gamma^{(\mathrm{eq})}}{\omega_{r}^{(\mathrm{eq})}}
\end{equation}
and one can see that $\gamma^{(\mathrm{eq})}$ acts as a Lorentz factor
which relates the elapse of proper time over one cycle of equatorial
motion with the corresponding period of coordinate time.

\subsection{Keplerian orbits}

In the Newtonian limit, the constants of motion asymptotically
approach the values of Keplerian orbits which are given by
equations~(\ref{eq:const_of_motion_kepl}) in
\ref{sct:const_of_motion}. Since the Keplerian values of
energy and angular momentum, naturally, are independent of the spin of
the black hole, the problem reduces to that one of motion in
Schwarzschild spacetime and we can set $z_{-}=0$.  Thus, we obtain the
the asymptotic expressions
\begin{eqnarray}
  \eqalign{
    F(\chi) \simeq\,1 + \mathrm{O}(p^{-1}), \qquad &
    G(\chi) \simeq\,p^{1/2} + \mathrm{O}(p^{-1/2}), \\
    H(\chi) \simeq\,1 + \mathrm{O}(p^{-1}), \qquad &
    J(\chi) \simeq\,(1-e^{2})^{2}p^{-1} + \mathrm{O}(p^{-2})
  }
\end{eqnarray}
which lead to
\begin{equation}
\fl \tilde{W} \simeq\,\tilde{Y}, \qquad
    \tilde{X} \simeq\,\frac{\pi p^{1/2}}{1-e^{2}}, \qquad
    \tilde{Y} \simeq\,\frac{\pi p}{(1-e^{2})\Omega_{\mathrm{K}}}, \qquad
    \tilde{Z} \simeq\,p^{1/2}\tilde{X},
\end{equation}
where $\Omega_{\mathrm{K}}$ is the Keplerian frequency given by
equation~(\ref{eq:kepler}).

Substitution of the above asymptotic expressions into 
equations~(\ref{eq:freq_r})--(\ref{eq:freq_phi}) and~(\ref{eq:gamma})
results in 
\begin{eqnarray}
    \gamma \simeq 1, \qquad &
    \tilde{\omega}_{r} \simeq \tilde{\omega}_{\theta} \simeq 
    \tilde{\omega}_{\phi} \simeq \Omega_{\mathrm{K}}.
\end{eqnarray}
The three components of motion therefore become commensurate in the
Newtonian limit and the corresponding frequencies asymptotically
degenerate. In the nearly Newtonian regime, where the three
frequencies are close to $\Omega_{\mathrm{K}}$ yet slightly different
from each other, we expect that there will be a minute perihelion
drift (due to the small difference between $\omega_{\phi}$ and
$\omega_{r}$) as well as a slow precession of the orbital plane (due
to the difference between $\omega_{\phi}$ and $\omega_{\theta}$). In
order to calculate this precession, an approximation of the
fundamental frequencies including terms of next lower order in $p$
would be required. However, this is left for future work.

\begin{figure}[hbf]
  \begin{center}
    \epsfig{file=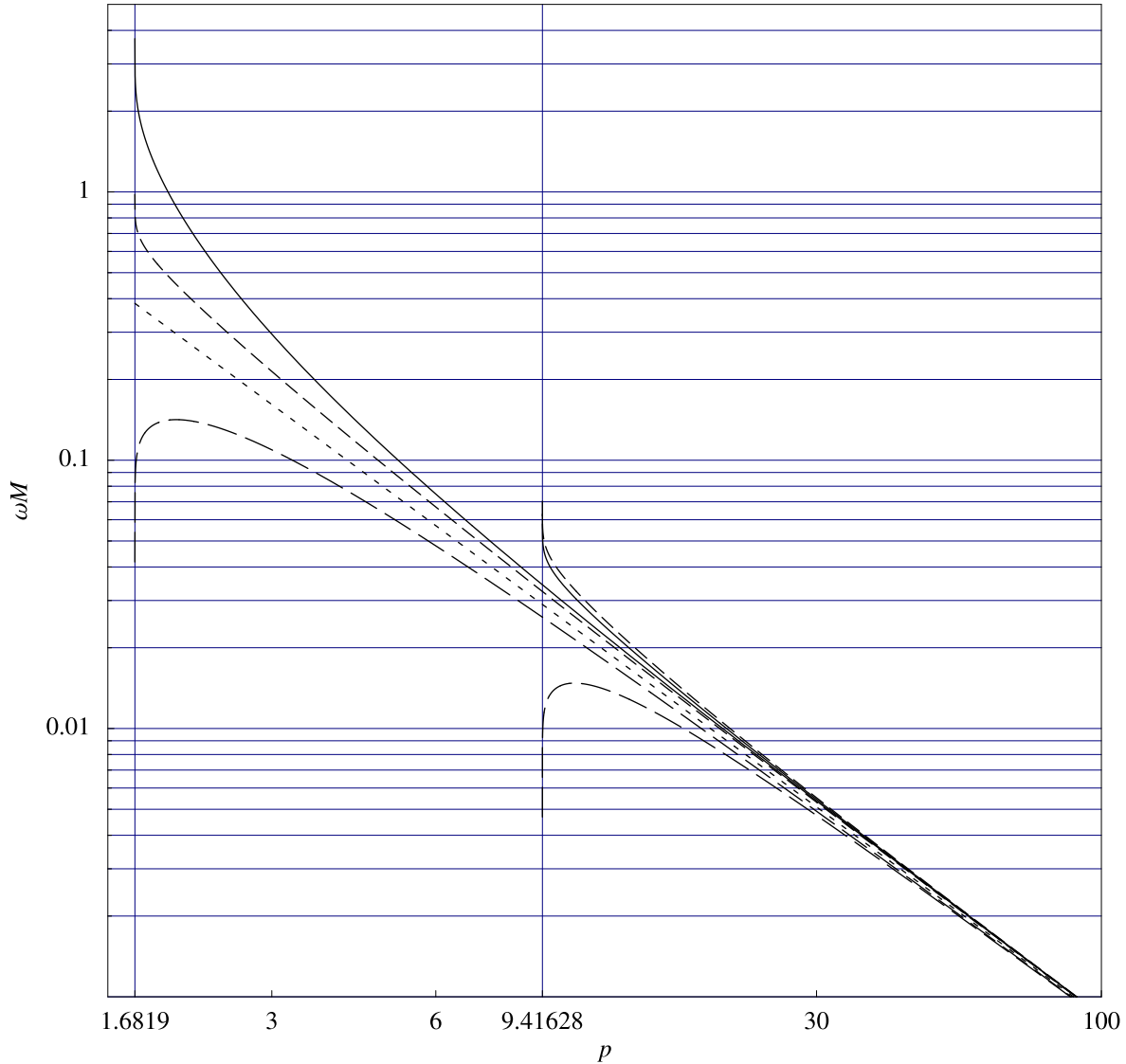,height=0.95\linewidth,clip}
    \caption{Dimensionless fundamental frequencies $\tilde{\omega}_{r}$,
      $\tilde{\omega}_{\theta}$ and $\tilde{\omega}_{\phi}$ as
      functions of $p$ for orbits of eccentricity $e=1/3$ and minimal
      inclination $\theta_{-}=\pi/3$ to the spin axis of the black
      hole (long-dashed lines: $\tilde{\omega}_{r}$, short-dashed
      lines: $\tilde{\omega}_{\theta}$, solid lines:
      $\tilde{\omega}_{\phi}$). The set of lines diverging towards
      $p\thickapprox 9.416$ corresponds to retrograde orbits (lower
      binding energy, smaller angular momentum), whereas the other set
      of lines which diverge towards $p\thickapprox 1.682$ are the
      frequencies of prograde orbits (higher binding energy, larger
      angular momentum). The two vertical grid lines mark
      the locations of the marginally stable prograde and
      retrograde orbits. The widely spaced dashed straight line is the
      "asymptotic Keplerian branch" which is given by the power law
      $\Omega_{\mathrm{K}}=(16\sqrt{2}/27)p^{-3/2}$.}
    \label{fg:freq_p}
  \end{center}
\end{figure}

\subsection{Numerical case study of a rapidly rotating black hole}

Let us now consider bound orbits around a nearly extreme Kerr black
hole of spin $\tilde{a}=0.998$.\footnote{ The chosen value of the spin
  is presumably a critical value as black holes which
  are accreting material tend to be buffered against further increase
  of spin if $\tilde{a}\approx 0.998$~\cite{Thorne74}.  } Such a black
hole is particularly interesting because effects due to spacetime
dragging caused by the high angular momentum of the black hole are
strong and stable orbits exist even very close to the horizon (which
is located at $r=r_{\mathrm{H}}\thickapprox 1.0632 M$).  Subsequently,
we will explore the parameter space in all three dimensions and
analyse the behaviour of the fundamental frequencies as each of the
orbital parameters is varied. The numerical evaluation of
equations~(\ref{eq:freq_r})--(\ref{eq:freq_phi}) and a general
algorithm to calculate the constants of motion $E$, $L_{z}$ and $Q$
for arbitrary orbital parameters $p$, $e$ and $\theta_{-}$ was
implemented in MATHEMATICA.\footnote{ The procedures
  were tested, for instance, by comparison of large numerical samples
  to the results of Wilkins~\cite{Wilkins72} for circular orbits
  around an extreme Kerr black hole and those of Cutler
  \etal~\cite{CutKenn94} for elliptical orbits in Schwarzschild
  spacetime. They are available from the author on request.  }

\begin{figure}[htb]
  \begin{center}
    \epsfig{file=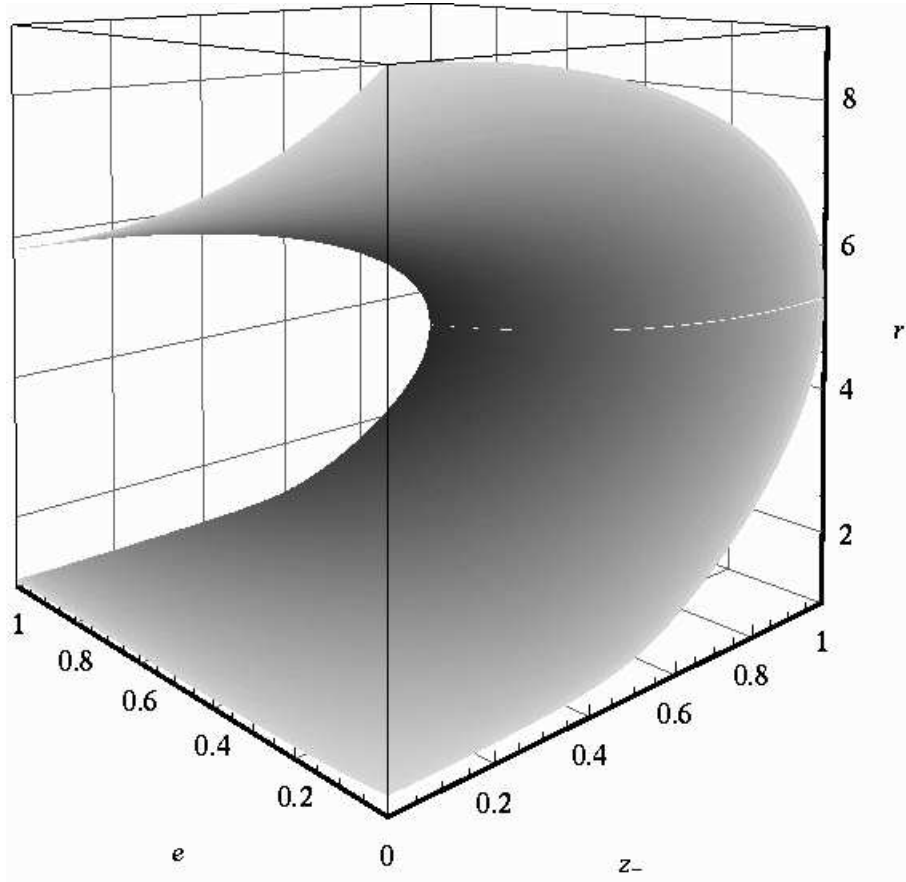,height=0.75\linewidth,clip}
    \caption{Separatrix of a Kerr black hole of spin $\tilde{a}=0.998$. The two 
      surfaces specify the smallest radial distance $r$ which can be
      reached by particles moving in a stable orbit of given
      eccentricity $e$ and inclination parameter $z_{-}$ without
      plunging into the black hole (a high-quality version of this
      graph can be obtained from the author).}
    \label{fg:sptrx}
  \end{center}
\end{figure}

To begin with, we choose $e=1/3$ (in this case the aphelion radius is
twice the perihelion radius) and $\theta_{-}=\pi/3$ (corresponding to
$30^{\circ}$ of maximum altitude from the equatorial plane). The
fundamental frequencies of the prograde and retrograde orbits as
functions of $p$ are shown in figure~\ref{fg:freq_p}. Basically, one
can see that the frequencies $\tilde{\omega}_{r}$,
$\tilde{\omega}_{\phi}$ and $\tilde{\omega}_{\theta}$ increasingly
deviate from each other as the distance from the black hole becomes
smaller. For increasing $p$, on the other hand, the fundamental
frequencies tend to degenerate and, in fact, they converge towards the
Keplerian frequency $\Omega_{\mathrm{K}}=(16\sqrt{2}/27)p^{-3/2}$ for
$e=1/3$. Thus, we call the Keplerian power law for given
eccentricity the \emph{asymptotic Keplerian branch} of the fundamental
frequencies.

\begin{figure}[hbf]
  \begin{center}
    \epsfig{file=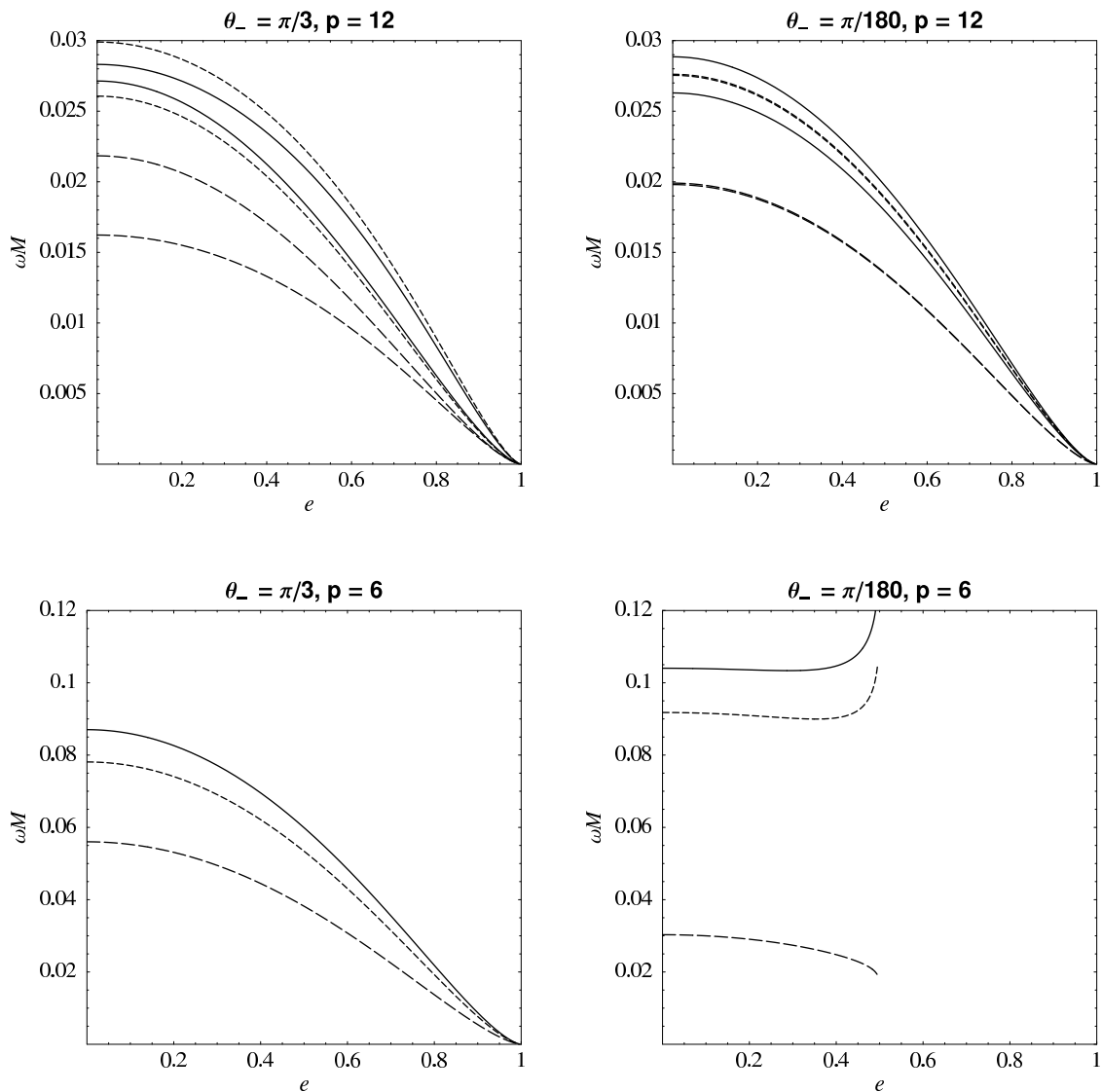, width=0.95\linewidth}
    \caption{Dimensionless fundamental frequencies $\tilde{\omega}_{r}$,
      $\tilde{\omega}_{\theta}$ and $\tilde{\omega}_{\phi}$ as
      functions of the eccentricity $e$ for different values of $p$
      and $\theta_{-}$ (long-dashed lines: $\tilde{\omega}_{r}$,
      short-dashed lines: $\tilde{\omega}_{\theta}$, solid lines:
      $\tilde{\omega}_{\phi}$; the frequencies of retrograde orbits
      enclose those of prograde orbits). In the bottom panels, only
      the frequencies of prograde orbits are plotted since there are
      no stable retrograde orbits if $p=6$.}
    \label{fg:freq_e}
  \end{center}
\end{figure}

The values of $p$, for which the frequencies of, respectively,
prograde and retrograde orbits diverge, are located exactly where the
respective orbits become \emph{marginally stable}.  In this case,
stable prograde orbits only exist if
$p>p_{\mathrm{ms}}^{(\mathrm{p})}\thickapprox 1.682$ and stable
retrograde orbits only if
$p>p_{\mathrm{ms}}^{(\mathrm{r})}\thickapprox 9.416$.  Hence, the
corresponding minimal perihelion radii of stable prograde and
retrograde orbits with $e=1/3$ and $\theta_{-}=\pi/3$ are,
respectively, $r_{\mathrm{p}}^{(\mathrm{p})} \thickapprox 1.261 M$ and
$r_{\mathrm{p}}^{(\mathrm{r})}\thickapprox 7.062 M$.  Note that
$\tilde{\omega}_{r}\rightarrow 0$ and
$\tilde{\omega}_{\phi}\rightarrow\infty$ as $p\rightarrow
p_{\mathrm{ms}}$.  This is a consequence of the orbits becoming "zoom
whirls" which means that the particle moves from the apastron inwards
and then revolves many times along an almost circular spiral before it
actually reaches the periastron.\footnote{ Equatorial zoom whirl
  orbits have been recently investigated \cite{GlampKenn01}.  } In the
limit of a marginally stable orbit, the periastron is approached
asymptotically.  The elapse of proper time as the particle moves from
apastron to periastron, thus, ever increases as $p\rightarrow
p_{\mathrm{ms}}$ and the radial frequency approaches zero.  If the
particle represents an astrophysical object, however, radiation
reaction will finally act on a time scale that is comparable to the
dynamical time scale of radial motion ($\sim 2\pi\gamma/\omega_{r}$)
and, as a consequence, it will drive the particle within finite time
over the stability threshold so that it plunges into the black hole.

\begin{figure}[hbf]
  \begin{center}
    \epsfig{file=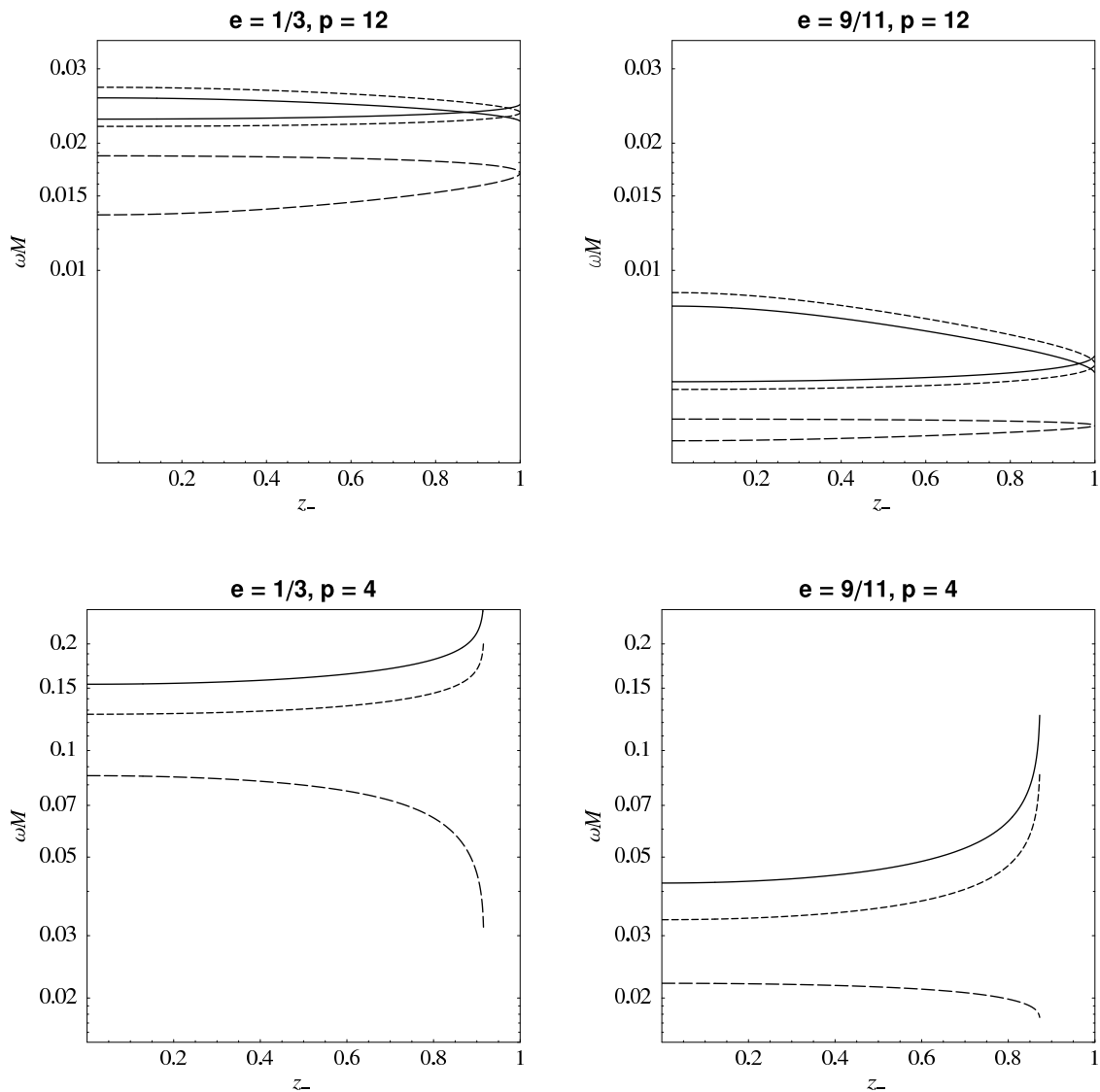, width=0.95\linewidth}
    \caption{Dimensionless fundamental frequencies $\tilde{\omega}_{r}$,
      $\tilde{\omega}_{\theta}$ and $\tilde{\omega}_{\phi}$ as
      functions of the minimal inclination to the spin axis of the
      black hole, $\theta_{-}$, for different values of $e$ and $p$.
      (long-dashed lines: $\tilde{\omega}_{r}$, short-dashed lines:
      $\tilde{\omega}_{\theta}$, solid lines: $\tilde{\omega}_{\phi}$;
      the frequencies of retrograde orbits enclose those of prograde
      orbits). In the bottom panels, only the frequencies of prograde
      orbits are plotted since stable retrograde orbits with $p=4$ do
      not exist at all. Even the prograde orbits are only stable if
      the altitude from the equatorial plane is lower than roughly
      $3\pi/8$ for $e=1/3$ and about $\pi/3$ for $e=9/11$.}
    \label{fg:freq_z}
  \end{center}
\end{figure}

In parameter space, the locus of all points corresponding to
marginally stable orbits, $p = p_{\mathrm{ms}}(e,z_{-})$, is called
the \emph{separatrix}. A plot of the separatrix for $\tilde{a}=0.998$
is shown in figure~\ref{fg:sptrx}, where the perihelion radius
$r_{\mathrm{p}}=p/(1+e)$ is used as parameter in place of $p$. There
are two sheets, one corresponding to prograde marginally stable orbits
and the other one corresponding to retrograde marginally stable
orbits.  The two sheets smoothly join at $z=1$, i.e., where the polar
orbits with $\tilde{L}_{z}=0$ are located.  As $z_{-}=\cos\theta_{-}$
becomes smaller, however, there is an increasing space between the two
sheets, in which only prograde but no retrograde orbits reside.  As
the eccentricity increases, both prograde and retrograde orbits exist
closer to the horizon due to the higher momentum of the particle at
the periastron as compared to circular orbits of the same radius. In
particular, there are prograde orbits of eccentricity $e\gtrsim 0.5$
with $z\lesssim 0.5$ for which the periastron is very close to the
horizon (within $\delta\tilde{r}\sim 0.1$).

\begin{figure}[hbf]
  \begin{center}
    \epsfig{file=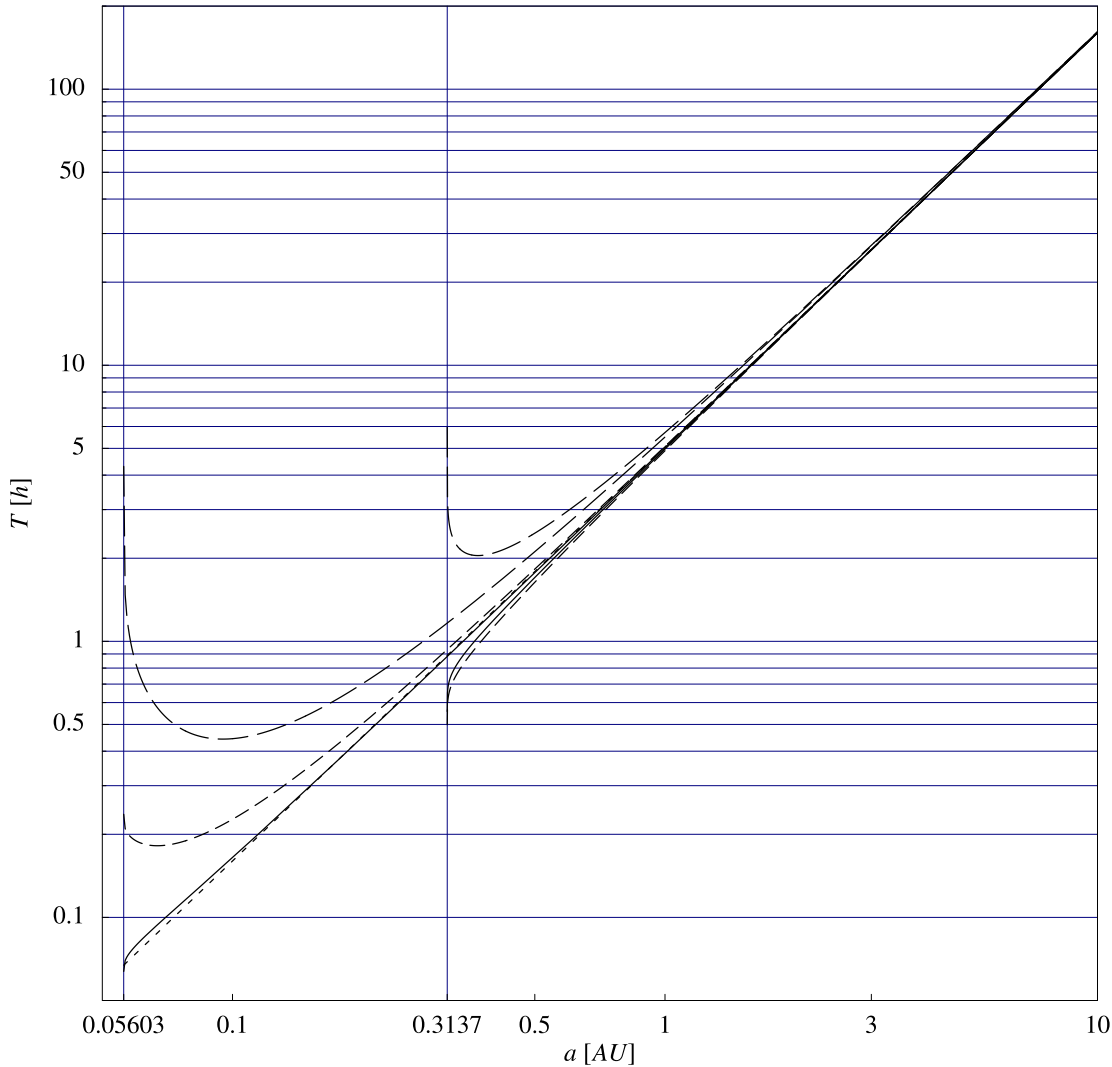,height=0.9\linewidth,clip}
    \caption{Scales of coordinate time $T_{r}$,
      $T_{\theta}$ and $T_{\phi}$ in hours as functions of the
      semi-major axis $a=p M/(1 - e^{2})$ in astronomical units
      ($1\,\mathrm{AU} = 1.496\cdot 10^{13}\,\mathrm{cm}$) for a
      central black hole of mass $M=3\cdot10^{6}M_{\mathrm{sol}}$
      (long-dashed lines: $T_{r}$, short-dashed lines: $T_{\theta}$,
      solid lines: $T_{\phi}$). As in figure~\ref{fg:freq_p}, the
      eccentricity of the orbits is $e=1/3$ and the inclination
      parameter $\theta_{-}=\pi/3$. The set of lines diverging towards
      $a\thickapprox 0.3137\,\mathrm{AU}$ corresponds to retrograde
      orbits, whereas the other set of lines which diverge towards
      $a\thickapprox 0.0560\,\mathrm{AU}$ are the frequencies of
      prograde orbits. The two vertical grid lines mark
      the locations of the marginally stable prograde and retrograde
      orbits. The widely spaced dashed straight line is the Keplerian
      orbital period given by $P_{\mathrm{K}}^{2}=4\pi^{2} a^{3}/G M$
      (in physical units).}
    \label{fg:time_p}
  \end{center}
\end{figure}

Figure~\ref{fg:freq_e} shows the change of the fundamental frequencies
with eccentricity for various choices of $\theta_{-}$ and $p$. The
main feature of these plots is the convergence of the frequencies to a
single value as $e\rightarrow 1$.  This simply reflects the fact that
orbits of eccentricity very close to unity are almost unbound and the
apastron is located far away from the region of strong gravity.  In
consequence, the particle is mostly moving in the nearly Newtonian
regions of spacetime and the fundamental frequencies tend to
degenerate. In the plot showing the frequencies of nearly polar orbits
with $\theta_{-}=\pi/180$ and $p=6$, the frequencies terminate just
below $e=0.5$ as prograde orbits of higher eccentricity do not exist
if $z\sim 1$ (note that the value of the perihelion radius depends on
$e$ according to $\tilde{r}_{\mathrm{p}}=6/(1+e)$ when comparing to
the plot of the separatrix in figure~\ref{fg:sptrx}).

The dependence of the fundamental frequencies on the inclination
parameter $\theta_{-}$ (which is the smallest angle of inclination to
the spin axis of the black hole that can be reached by the particle)
is shown in figure~\ref{fg:freq_z} for different values of $e$ and
$p$. Apparently, both the radial and polar frequencies of prograde and
retrograde orbits converge towards each other as
$\theta_{-}\rightarrow 0$, i.e., in the polar limit. On the other
hand, there is a crossing-over of the azimuthal frequencies at some
finite value of $\theta_{-}$.  As a consequence, the azimuthal motion
of nearly polar retrograde orbits is \emph{faster} than that of the
corresponding prograde orbits and, thus, it seems that nearly polar
orbits are markedly different from nearly equatorial orbits. We have
to keep in mind, however, that the revolution of the particle around
the black hole in the case $\theta_{-}\sim 0$ is rather specified by
the polar frequency $\omega_{\theta}$, whereas $\omega_{\phi}$
determines an azimuthal precession around the spin axis of the black
hole.

Finally, in figure~\ref{fg:time_p}, the scales of coordinate
time, $T_{r}$, $T_{\theta}$ and $T_{\phi}$, defined by
equation~\ref{eq:coord_time_scl} are shown in physical units as
functions of the semi-major axis $a$ for orbits of eccentricity
$e=1/3$ and inclination parameter $\theta_{-}=\pi/3$. The mass of the
black hole is chosen to be $M=3\cdot10^{6}M_{\mathrm{sol}}$ which is
about the mass determined for the black hole in the centre of our
Galaxy~\cite{HoKorm00}.  It is interesting to note that for virtually
all prograde orbits the time scale of polar motion $T_{\phi}$ is
very close to the the Keplerian orbital period
$P_{\mathrm{K}}=2\pi/\Omega_{\mathrm{K}}$. However, this is only accidentally
the case for the chosen set of parameters, as for lower eccentricity
$T_{\phi}$ is above $P_{\mathrm{K}}$, whereas it tends to become
significantly smaller than $P_{\mathrm{K}}$ at higher eccentricity.

\section{Conclusion}

The investigation of bound geodesic orbits in Kerr spacetime presented
in this article clearly illustrates that the properties of these orbits in the
regime of strong gravity are profoundly different from Keplerian
orbits in the Newtonian regime. In particular, there are three
different frequencies of orbital motion, $\omega_{r}$,
$\omega_{\theta}$ and $\omega_{\phi}$, which increasingly deviate from
the Keplerian frequency as the size of the orbit decreases and finally
approaches the limit of a marginally stable orbit. This is a
consequence of the radial, polar and azimuthal components of motion
being \emph{incommensurate} in Kerr spacetime. As a consequence, the
shapes of orbits in the vicinity of a Kerr black hole are
exceedingly complex.

The explicit formulas for the fundamental frequencies $\omega_{r}$,
$\omega_{\theta}$ and $\omega_{\phi}$ presented in
Section~\ref{sct:fund_freq} are exact in the test particle limit and
useful approximations if radiation reaction is \emph{adiabatic}, i.e.,
if the time scales associated with the frequencies are small compared
to the time scale of radiation reaction. Based on the constant of
motion $\gamma$ which is associated with a timelike generalised
coordinate, we have proposed a conversion of the invariant fundamental
frequencies into coordinate-time scales, $T_{r}$, $T_{\phi}$ and
$T_{\theta}$, which should be relevant to an observer at spatial
infinity and, thus, gravitational wave astronomy. Once these time
scales are established, one can compute \emph{instantaneous}
gravitational wave forms in the adiabatic approximation. In this
context, the term ``instantaneous'' means an interval of time, over
which no significant changes of the constants of motion are
accumulated due to radiation reaction.  From the resulting spectrum,
an estimation of the power emitted in the frequency band of a detector
could be made, without making use of an advanced theory of radiation
reaction.

\section*{Acknowledgements}

I thank C.\ Glampedakis for many discussions and making me aware of
the problem concerning the dynamical time scales of geodesic orbits in
Kerr spacetime and also B.\ S.\ Sathyaprakash and L.\ P.\ Grishchuk
for their support and useful criticism. This work was funded by a
\emph{Wilhelm-Macke mobility scheme award} of the
Johannes-Kepler-University Linz, Austria, as part of my postgraduate
studies at Cardiff University.

\appendix

\section{Determination of the fundamental frequencies}

\label{sct:determ_freq}

Let
$P_{\beta}^{(\mathrm{aa})}=f_{\beta}^{(\mathrm{aa})}(-\mu^{2}/2,E,L_{z},Q)$
be the momenta given by $P_{0}^{(\mathrm{aa})}=p_{t}=-E$ and
$P_{k}^{(\mathrm{aa})}=J_{k}$, where the $J_{k}$ are the action
variables defined by
equations~(\ref{eq:action_var_r})--(\ref{eq:action_var_phi}).  If we
denote the Jacobian matrix of $f$ by $\mathsfsl{D}f$, then, by the
theorem on implicit functions,
$\mathsfsl{D}f\cdot\mathsfsl{D}(f^{-1})=
\mathsfsl{D}f\cdot(\mathsfsl{D}f)^{-1}=\mathsfsl{I}$, provided that
$f$ is non-zero and the Jacobian does not vanish~\cite{Hille}.  Since
$-\mu^2/2$ is the invariant value of the Hamiltonian, we can
substitute $-\mu^2/2=H^{(\mathrm{aa})}(-E,J_{k})$, where
$H^{(\mathrm{aa})}$ is the Hamiltonian in the action-angle variable
representation. For brevity, let us subsequently use the symbol $H$ to
denote the Hamiltonian in that representation. Moreover, two rows of
the Jacobian matrix are trivial due to the identities
$P_{0}^{(\mathrm{aa})}=-E$ and $J_{\phi}=L_{z}$. Thus, the equation
$\mathsfsl{D}f\cdot\mathsfsl{D}(f^{-1})=\mathsfsl{I}$ reads
\begin{equation}
  \left(\begin{array}{llll}
    0 & -1 & 0 & 0 \\
    \frac{\partial J_{r}}{\partial H} & 
    \frac{\partial J_{r}}{\partial E} &
    \frac{\partial J_{r}}{\partial L_{z}} & 
    \frac{\partial J_{r}}{\partial Q} \\
    \frac{\partial J_{\theta}}{\partial H} & 
    \frac{\partial J_{\theta}}{\partial E} &
    \frac{\partial J_{\theta}}{\partial L_{z}} & 
    \frac{\partial J_{\theta}}{\partial Q} \\
    0 & 0 & 1 & 0
  \end{array}\right)\cdot
  \left(\begin{array}{llll}
    -\frac{\partial H}{\partial E} &
    \frac{\partial H}{\partial J_{r}} &
    \frac{\partial H}{\partial J_{\theta}} &
    \frac{\partial H}{\partial J_{\phi}} \\
    -1 & 0 & 0 & 0 \\
    0 & 0 & 0 & 1 \\
    -\frac{\partial Q}{\partial E} &
    \frac{\partial Q}{\partial J_{r}} &
    \frac{\partial Q}{\partial J_{\theta}} &
    \frac{\partial Q}{\partial J_{\phi}} \\
   \end{array}\right) = \mathsfsl{I}.
\end{equation}
The above matrix equation can be split into four non-trivial sets of
linear equations in the eight unknowns $-\frac{\partial H}{\partial
  E}$, $\frac{\partial H}{\partial J_{k}}$, $-\frac{\partial
  Q}{\partial E}$ and $\frac{\partial Q}{\partial J_{k}}$:
\begin{eqnarray}
  -&\mathsfsl{A}\cdot\frac{\partial}{\partial E}
  \left(\begin{array}{llll}  H \\ Q \end{array}\right) =
  \left(\begin{array}{llll} 
    2 \int_{r_{1}}^{r_{2}}\frac{\rmd r}{\sqrt{R}}
    \left[\frac{(r^{2}+a^{2}) P}{\Delta} + a(L_{z}-a E)\right] \\
    2 a^{2} E\int_{\theta_{-}}^{\pi/2}
    \frac{\cos^{2}\theta\rmd \theta}{\sqrt{\Theta}}
  \end{array}\right) \\
  \label{eq:syst_act_r}
  &\mathsfsl{A}\cdot\frac{\partial}{\partial J_{r}}
  \left(\begin{array}{llll}  H \\ Q \end{array}\right) =
  \left(\begin{array}{llll} 2\pi \\ 0 \end{array}\right), \\
  &\mathsfsl{A}\cdot\frac{\partial}{\partial J_{\theta}}
  \left(\begin{array}{llll}  H \\ Q \end{array}\right) =
  \left(\begin{array}{llll} 0 \\ \frac{\pi}{2} \end{array}\right), \\
  &\mathsfsl{A}\cdot\frac{\partial}{\partial J_{\phi}}
  \left(\begin{array}{llll}  H \\ Q \end{array}\right) =
  \left(\begin{array}{llll}
    2 \int_{r_{1}}^{r_{2}}\frac{\rmd r}{\sqrt{R}}
    \left[\frac{a P}{\Delta}+(L_{z}-a E)\right]  \\
    2 L_{z}\int_{\theta_{-}}^{\pi/2}
    \frac{\cot^{2}\theta\rmd \theta}{\sqrt{\Theta}}
  \end{array}\right)
\end{eqnarray}
where $P=(r^{2}+a^{2})E - a L_{z}$, and the coefficient matrix
$\mathsfsl{A}$ is given by
\begin{equation}
  \mathsfsl{A} = 
  \left(\begin{array}{llll} 
    2\int_{r_{1}}^{r_{2}}\frac{r^{2}\rmd r}{\sqrt{R}} & 
    -\int_{r_{1}}^{r_{2}}\frac{\rmd r}{\sqrt{R}} \\ 
    2 a^{2}\int_{\theta_{-}}^{\pi/2}
    \frac{\cos^{2}\theta\,\rmd \theta}{\sqrt{\Theta}} &
    \int_{\theta_{-}}^{\pi/2}\frac{\rmd \theta}{\sqrt{\Theta}}.
  \end{array}\right)
\end{equation}

Using the definitions~(\ref{eq:rad_integr_x})--(\ref{eq:rad_integr_z})
and~(\ref{eq:rad_integr_w}) for the radial integrals and the identities
\begin{eqnarray}
  \int_{\theta_{-}}^{\pi/2}\frac{\rmd \theta}{\sqrt{\Theta}} & = 
  \frac{1}{\beta z_{+}}K(k), \\
  \label{eq:angl_intgr2}
  \int_{\theta_{-}}^{\pi/2}\frac{\cos^{2}\theta}{\sqrt{\Theta}}\,\rmd \theta & = 
  \frac{z_{+}}{\beta}[K(k)-E(k)], \\
  \label{eq:angl_intgr3}
  \int_{\theta_{-}}^{\pi/2}\frac{\cot^{2}\theta}{\sqrt{\Theta}}\,\rmd \theta & = 
  \frac{z_{+}}{\beta}[\Pi(z_{-}^{2},k)-K(k)],
\end{eqnarray}
where the elliptical integrals $K(k)$, $E(k)$ and $\Pi(-z_{-}^{2},k)$
are defined by
equations~(\ref{eq:ell_integr_k})--(\ref{eq:ell_integr_pi}), the
solutions of the above systems of equations for $-\frac{\partial
  H}{\partial E}$ and $\frac{\partial H}{\partial J_{k}}$ are given by
\begin{eqnarray}
  \label{eq:h_derv_p0}
  -&\frac{\partial H}{\partial E} =
  \frac{K(k)W(r_{1},r_{2}) + 
        a^{2}z_{+}^{2}E\left[K(k)-E(k)\right]X(r_{1},r_{2})
        }{K(k)Y(r_{1},r_{2}) +
          a^{2}z_{+}^{2}\left[K(k)-E(k)\right]X(r_{1},r_{2})}, \\
  \label{eq:h_derv_jr}
  &\frac{\partial H}{\partial J_{r}} =
  \frac{\pi K(k)}{K(k)Y(r_{1},r_{2}) + a^{2}z_{+}^{2}\left[K(k)-E(k)\right]
                  X(r_{1},r_{2})}, \\
  \label{eq:h_derv_jtheta}
  &\frac{\partial H}{\partial J_{\theta}} =
  \frac{\pi\beta z_{+}X(r_{1},r_{2})}{2\{K(k)Y(r_{1},r_{2}) +
    a^{2}z_{+}^{2}\left[K(k)-E(k)\right]X(r_{1},r_{2})\}}, \\
  \label{eq:h_derv_jphi}
  &\frac{\partial H}{\partial J_{\phi}} =
  \frac{K(k)Z(r_{1},r_{2}) + L_{z}[\Pi(z_{-}^{2},k)-K(k)]X(r_{1},r_{2})
        }{K(k)Y(r_{1},r_{2}) + a^{2}z_{+}^{2}[K(k)-E(k)]X(r_{1},r_{2})}.
\end{eqnarray}

The validity of the above solution depends on the Jacobian matrix
$\mathsfsl{D}f$ being non-singular. In fact, $\det\mathsfsl{D}f\neq 0$
if and only if
\begin{equation}
\fl \det\mathsfsl{D}(f^{-1}) =
  - \frac{\partial Q}{\partial J_{r}}\cdot\frac{\partial H}{\partial J_{r}} =
  2a^{2}z_{+}^{2}\left[1-\frac{E(k)}{K(k)}\right]
  \left(\frac{\partial H}{\partial J_{r}}\right)^{2}
  \neq 0,
\end{equation}
as is easily shown by using the second equation of
system~(\ref{eq:syst_act_r}).  The above condition is fulfilled
whenever $k>0$, i.e., if the orbit is non-equatorial.  Since $Q=0$ for
equatorial orbits, this is immediately clear from the above systems of
equations as well. Nevertheless, the frequencies of equatorial orbits
can be calculated by taking the limit $k\rightarrow 0$ of the general
expressions for $k>0$ and, as demonstrated in Section~\ref{sct:equat},
one finds agreement with the results obtained from a direct
calculation based on the equations of motion.

\section{Calculation of the constants of motion}

\label{sct:const_of_motion}

In order to find the constants of motion $E$, $L_{z}$ and $Q$ for
given orbital parameters $p$, $e$ and $\theta_{-}$, one has to solve
the following set of equations:
\begin{eqnarray}
  \frac{\rmd r}{\rmd \tau}      & = 0 \qquad \Longleftrightarrow
  \qquad R(r)      & = 0, \\
  \frac{\rmd \theta}{\rmd \tau} & = 0 \qquad \Longleftrightarrow
  \qquad \Theta(\theta) & = 0,
\end{eqnarray}
where $R(r)$ and $\Theta(\theta)$ are given by
equations~(\ref{eq:capt_r}) and~(\ref{eq:capt_theta}). The roots of
these equations correspond to the turning points of, respectively,
radial and polar motion, i.e.,we can substitute $r=r_{\mathrm{p}}=p
M/(1+e)$ or $r=r_{\mathrm{a}}=p M/(1-e)$ in the first equation and
$\theta=\theta_{-}$ in the second equation.  Thus, a system of three
equations is obtained which can be solved for the three unknown
constants of motion, provided that $e\neq 0$.  In the case $e=0$, on
the other hand, we have got to make use of the supplemental constraint
$R'(r_{0})=0$, in addition to $R(r_{0})=0$ and $\Theta(\theta_{-})=0$,
because circular orbits occur if both $R(r)$ and its radial gradient
$R'(r)$ vanish at the same point, $r=r_{0}$.

Introducing the dimensionless quantities
\begin{equation}
  \tilde{a} = \frac{a}{M}, \quad 
  \tilde{E} = \frac{E}{\mu}, \quad
  \tilde{L}_{z} = \frac{L_{z}}{\mu M}, \quad 
  \tilde{Q} = \frac{Q}{\mu^{2} M^{2}},
\end{equation}
we can actually set up a completely scale-invariant formulation of the
problem. Hence, it is sufficient to calculate $\tilde{E}$,
$\tilde{L}_{z}$ and $\tilde{Q}$ only once for given orbital
parameters. The corresponding physical values for a particular black
hole of mass $M$ and a particle of mass $\mu$ can then be inferred
from the above definitions.

To begin with, we re-arrange $\Theta(\theta_{-})=0$ in order to
express Carter's constant $\tilde{Q}$ in terms of $\theta_{-}$,
$\tilde{E}$ and $\tilde{L}_{z}$,
\begin{equation}
  \tilde{Q} = z_{-}^{2}\left[\tilde{a}^{2}(1-\tilde{E}^{2}) + 
    \frac{\tilde{L}_{z}^{2}}{1-z_{-}^{2}}\right],
\end{equation}
where $z_{-}=\cos\theta_{-}$. If this expression is substituted for
$\tilde{Q}$ in the equation
$\tilde{R}(\tilde{r})=\mu^{-2}M^{-4}R(r)=0$, we obtain
\begin{equation}
  \tilde{R}(\tilde{r}) = f(r)\tilde{E}^{2} - 2 g(r)\tilde{E}\tilde{L}_{z} -
                         h(r) \tilde{L}_{z}^{2} - d(r),
\end{equation}
where   
\begin{eqnarray}
  f(r) & = \tilde{r}^{4} + \tilde{a}^{2}  
           \left[\tilde{r}(\tilde{r}+2)+z_{-}^{2}\tilde{\Delta}\right], \\
  g(r) & = 2\tilde{a}\tilde{r}, \\                                     
  h(r) & = \tilde{r}(\tilde{r}-2) + \frac{z_{-}^{2}}{1-z_{-}^{2}}\tilde{\Delta}, \\
  d(r) & = (\tilde{r}^{2} + \tilde{a}^{2}z_{-}^{2})\tilde{\Delta},
\end{eqnarray}
and $\tilde{\Delta}=\tilde{r}^{2}-2\tilde{r}+\tilde{a}^{2}$.
Substitution of $\tilde{Q}$ in $\tilde{R}'(r)=0$ yields
\begin{equation}
  \frac{\rmd \tilde{R}}{\rmd \tilde{r}} = 
  f'(r)\tilde{E}^{2} - 2 g'(r)\tilde{E}\tilde{L}_{z} - 
  h'(r)\tilde{L}_{z}^{2} - d'(r),
\end{equation}
where
\begin{eqnarray}
  f'(r) & = 4\tilde{r}^{3} + 2\tilde{a}^{2}       
            \left[(1+z_{-}^{2})\tilde{r}+(1-z_{-}^{2})\right], \\
  g'(r) & = 2\tilde{a}, \\                                     
  h'(r) & = \frac{2(\tilde{r}-1)}{1-\tilde{z}_{-}^{2}}, \\
  d'(r) & = 2\left(2\tilde{r}-3\right)\tilde{r}^{2} +
            2\tilde{a}^{2}\left[(1+z_{-}^{2})\tilde{r}-z_{-}^{2}\right].
\end{eqnarray}

Thus, we shall define the following coefficients in terms of the orbital
parameters:
\begin{eqnarray}
  (f_{1}, g_{1}, h_{1}, d_{1}) & =
  \left\{\begin{array}{ll} 
    (f(\tilde{r}_{\mathrm{p}}), g(\tilde{r}_{\mathrm{p}}), 
     h(\tilde{r}_{\mathrm{p}}), d(\tilde{r}_{\mathrm{p}})) & 
    \mbox{if $e>0$}, \\
    (f(\tilde{r}_{0}), g(\tilde{r}_{0}), h(\tilde{r}_{0}), d(\tilde{r}_{0})) & 
    \mbox{if $e=0$},
  \end{array}\right. \\
  (f_{2}, g_{2}, h_{2}, d_{2}) & =
  \left\{\begin{array}{ll}
    (f(\tilde{r}_{\mathrm{a}}), g(\tilde{r}_{\mathrm{a}}), 
     h(\tilde{r}_{\mathrm{a}}), d(\tilde{r}_{\mathrm{a}})) & 
    \mbox{if $e>0$}, \\
    (f'(\tilde{r}_{0}), g'(\tilde{r}_{0}), h'(\tilde{r}_{0}),
     d'(\tilde{r}_{0})) & 
    \mbox{if $e=0$}.
  \end{array}\right.
\end{eqnarray}
Eliminating $\tilde{L}_{z}$ from the set of quadratic equations,
\begin{equation}
  \label{eq:energy_angmomt}
  \left.
    \begin{array}{ll}
      f_{1}\tilde{E}^{2} - 2 g_{1}\tilde{E}\tilde{L}_{z} - 
      h_{1}\tilde{L}_{z}^{2} - d_{1} & = 0, \\
      f_{2}\tilde{E}^{2} - 2 g_{2}\tilde{E}\tilde{L}_{z} - 
      h_{2}\tilde{L}_{z}^{2} - d_{2} & = 0, \\
    \end{array}
  \right\}
\end{equation}
a quadratic equation in $\tilde{E}^{2}$ is obtained,
\begin{equation}
  \label{eq:energy}
  (\rho^{2}+4\eta\sigma)\tilde{E}^{4} - 
  2(\kappa\rho+2\epsilon\sigma)\tilde{E}^{2} + \kappa^{2} = 0,
\end{equation}
where $\epsilon$, $\eta$, $\kappa$, $\rho$ and $\sigma$ are defined by
the following $2\times 2$ determinants:
\begin{eqnarray}
  \eqalign
  \kappa   = & 
  \left| \begin{array}{cc} 
      d_{1} & h_{1} \\ d_{2} & h_{2} 
  \end{array} \right|, \qquad
  \epsilon = & 
  \left| \begin{array}{cc} 
      d_{1} & g_{1} \\ d_{2} & g_{2} 
  \end{array} \right|, \\
  \rho     = & 
  \left| \begin{array}{cc} 
      f_{1} & h_{1} \\ f_{2} & h_{2} 
  \end{array} \right|, \qquad
  \eta     = & 
  \left| \begin{array}{cc} 
      f_{1} & g_{1} \\ f_{2} & g_{2} 
  \end{array} \right|, \\
  \sigma   = & 
  \left| \begin{array}{cc} 
      g_{1} & h_{1} \\ g_{2} & h_{2} 
  \end{array} \right|. &
\end{eqnarray}
The roots of equation~(\ref{eq:energy}) are given by
\begin{equation}
  \label{eq:energy_soln}
  \tilde{E}_{\pm}^{2} =
  \frac{\kappa\rho+2\epsilon\sigma \pm 
        2\sqrt{\sigma(\sigma\epsilon^{2}+\rho\epsilon\kappa-\eta\kappa^{2})}
        }{\rho^{2}+4\eta\sigma}.
\end{equation}
Actually, these roots can be calculated regardless of whether the
corresponding orbits are of the first or the second kind, i.e., if
they are stable or plunging. There is a simple criterion, however,
which can be used to determine the kind of the orbit: Let the
\emph{potential} of the radial component of motion be defined by
\begin{equation}
  \left(\frac{\rmd r}{\rmd \tau}\right)^{2} = \tilde{E}^{2} - \tilde{V}^{2}.
\end{equation}
According to equation~(\ref{eq:motion_r}),
$\tilde{V}=(\tilde{E}^{2}-\tilde{R}/\tilde{\Sigma}^{2})^{1/2}$ and the
radial domain of motion is subject to the constraint
$\tilde{V}\leq\tilde{E}$. Hence, the orbit is \emph{stable} if
\begin{equation}
  \frac{\partial\tilde{V}}{\partial\tilde{r}}(\tilde{r}_{\mathrm{p}}) < 0.
\end{equation}
If the gradient of the potential vanishes at $r=r_{\mathrm{p}}$, the
potential has a local maximum at the perihelion radius and, in that
case, the orbit is \emph{marginally stable}.

Substituting $\pm\sqrt{E_{\pm}^{2}}$ into the
system~(\ref{eq:energy_angmomt}) and solving for $L_{z}$, yields four
roots which satisfy both equations. Therefore, we obtain \emph{four}
possible solutions for the constants of motion with given orbital
parameters, namely,
$(-E^{(\mathrm{r})},-L_{z}^{(\mathrm{r})},Q^{(\mathrm{r})})$,
$(-E^{(\mathrm{p})},-L_{z}^{(\mathrm{p})},Q^{(\mathrm{p})})$,
$(E^{(\mathrm{p})},L_{z}^{(\mathrm{p})},Q^{(\mathrm{p})})$ and
$(E^{(\mathrm{r})},L_{z}^{(\mathrm{r})},$ $Q^{(\mathrm{r})})$, where
$E^{(\mathrm{p})}=E_{-}$ is the \emph{lower} and
$E^{(\mathrm{r})}=E_{+}$ the \emph{higher} energy given by
equation~(\ref{eq:energy_soln}). Numerical evaluation shows that
$L_{z}^{(\mathrm{p})}>L_{z}^{(\mathrm{r})}$. In the first case, the
particle has higher binding energy and co-revolves with the rotation
of the black hole. Such orbits are called \emph{prograde}. In the
second case, the particle has lower binding energy and usually
counter-revolves, except for some orbits close to the horizon of
rapidly rotating black holes. These orbits are called
\emph{retrograde}.  The solutions with negative energies correspond to
the respective motions under time reversal.

Finally, we consider the \emph{Newtonian} limits of the constants of
motion which are asymptotically approached as $p\rightarrow\infty$.
Using the formulas for the potential energy and angular momentum of
Keplerian orbits in geometric units~\cite{CarrOst},
\begin{equation}
  \tilde{U} = -\frac{1-e^{2}}{2p}, \qquad
  \tilde{L} = p^{1/2},
\end{equation}
and noting that
$L_{z}=L\cos(\pi/2-\theta_{-})=L\sin\theta_{-}=L(1-z_{-}^{2})^{1/2}$,
we obtain the following asymptotic expressions for the constants of
motion:\footnote{ The constraints
  $\tilde{R}(\tilde{r}_{\mathrm{p}})=0$ and
  $\tilde{R}(\tilde{r}_{\mathrm{a}})=0$ are fulfilled asymptotically
  in the highest power for non-circular orbits, and
  $\tilde{R}(\tilde{r}_{0})=0$ is satisfied down to
  $\mathrm{O}(\tilde{r}^{5/2})$ if $e=0$.  }
\begin{equation}
  \label{eq:const_of_motion_kepl}
  1-\tilde{E}^{2}   \simeq\, \frac{1-e^{2}}{p}, \qquad 
  \tilde{L}_{z}^{2} \simeq\, (1-z_{-}^{2})p, \qquad 
  \tilde{Q}         \simeq\, z_{-}^{2}p.
\end{equation}
Since $\tilde{Q} + \tilde{L}_{z}^{2} \simeq \tilde{L}^{2}$, Carter's
constant in the Newtonian limit is the squared angular momentum component
parallel to the equatorial plane.

\end{document}